\documentclass{aa}
\usepackage{natbib}
\usepackage{graphicx}
\usepackage{amssymb}

\bibpunct{(}{)}{;}{a}{}{,}

\begin{document}

\title{Fingerprints of the hierarchical building up of the structure on the gas kinematics of galaxies}

\author{M. E. De Rossi \inst{1,2,3}
        \and P. B. Tissera \inst{1,2}
        \and S. E. Pedrosa\inst{1,2}
        }

\offprints{M. E. De Rossi}

\institute{Consejo Nacional de Investigaciones  Cient\'{\i}ficas y T\'ecnicas, CONICET, Argentina\\
\email{derossi@iafe.uba.ar}
\and Instituto de Astronom\'{\i}a y F\'{\i}sica del Espacio, Casilla de Correos 67, Suc. 28, 1428, Ciudad Aut\'onoma de Buenos Aires, Argentina\\
\email{patricia@iafe.uba.ar,supe@iafe.uba.ar}
\and Facultad de Ciencias Exactas y Naturales, Universidad de Buenos Aires, Ciudad Aut\'onoma de Buenos Aires, Argentina\\
}

\date{Received / Accepted}

\abstract {Recent observational and theoretical works have suggested that the Tully-Fisher Relation might be generalised to include
dispersion-dominated systems by combining the rotation and dispersion velocity  in the definition of the kinematical
indicator. 
Mergers and interactions have been pointed out as responsible of driving turbulent and disordered
gas kinematics, which could generate Tully-Fisher Relation outliers.} 
{We intend to investigate the 
gas kinematics of galaxies by using a simulated sample which includes both, gas disc-dominated and spheroid-dominated systems.
We pay particular attention to the evolution of
the scatter of the Tully-Fisher Relation.
We also aim at determining the gas-phase velocity indicator
which better traces the potential well of the galaxy.}
{Cosmological hydrodynamical simulations which include a multiphase model and physically-motivated
Supernova feedback were performed in order to
follow the  evolution of galaxies as they are assembled. We analyse the gas kinematics 
of the surviving gas discs
to estimate all velocity indicators. 
} 
{
Both the baryonic and stellar Tully-Fisher relations for gas disc-dominated systems 
are tight while,
as more dispersion-dominated systems are included, the scatter increases.
We found a clear correlation between  $\sigma / V_{\rm rot}$ and morphology, with dispersion-dominated
systems exhibiting the larger values ($> 0.7$).
Mergers and interactions can  affect the rotation curves directly or indirectly 
inducing a scatter in the Tully-Fisher Relation larger than the simulated evolution since $z \sim 3$.
Kinematical indicators which
combine rotation velocity and dispersion velocity can reduce the scatter in the baryonic and the stellar mass-velocity relations. 
In particular, $s_{1.0} = ( V_{\rm rot}^2 + {\sigma}^2 )^{0.5} $ seems to be the best tracer of the circular velocity at
larger radii.
Our findings also show that the lowest scatter in both relations is
obtained if the velocity indicators are measured at the maximum of the rotation curve.
}
{In agreement with previous works, we found that the gas kinematics of galaxies
is significantly regulated by mergers and interactions, which 
play a key role in inducing gas accretion, outflows and starbursts. 
The joint action of these processes within a hierarchical $\Lambda$CDM Universe generates
a mean simulated Tully-Fisher Relation in good agreement with observations since $z \sim 3$ but with
a scatter depending on morphology. 
The rotation velocity  estimated at the
maximum of the gas rotation curve is found to be the best proxy for the potential well regardless of morphology.
}{}


\titlerunning{Gas kinematics of galaxies}
\authorrunning{De Rossi et al.}

\maketitle

\section{Introduction}
The Tully-Fisher Relation \citep[][hereafter TFR]{tully1977} is considered one of the most
fundamental scaling relations of disc galaxies as it links two important
properties: their stellar (sTFR) or baryonic (bTFR)
mass content and the depth of their potential well measured by the rotation velocity.
  Therefore, the study of the TFR
at different redshifts ($z$) could help to trace the dynamics of the gas and the star
formation activity within the dark matter haloes, and hence, to  provide constraints
for galaxy formation models \citep[e.g.][]{avila1998, mo1998, avila2008}.

The determination of the slope and zero point of the TFR at low 
\citep[e.g.][]{bell2001,
pizagno2007, meyer2008, gurovich2010} and intermediate and high redshifts 
\citep[e.g.][]{concelise2005, flores2006, atkinson2007, kassin2007,
puech2008, cresci2009, gnerucci2011}  has been the subject
of numerous works.
Although many authors suggest the existence of  evolution,  observational results
have not  converged yet, being also unclear  if the evolution is present in the zero point, the slope, or in both
\citep{vogt1996, vogt1997, nakamura2006}.
Recent observational results 
by \citet{miller2011} 
show evidence for a modest evolution of
the sTFR 
by $0.04 \pm 0.07$  from $z \sim 1$ to $z \sim 0.3$ 
in the sense that, at a given rotation velocity, galaxies have lower stellar masses in the
past.  
According to the results of \citet{cresci2009}, the sTFR seems 
to be already in place at $z \sim 2$ with a slope similar to the local one
but displaced towards lower stellar masses by $0.41 \pm 0.11$ dex.
These trends are in general good
agreement with predictions of  hydrodynamical simulations \citep[e.g.][]{portinari2007,derossi2010}.
At higher redshifts, it is not yet clear if the TFR is at place due to the high scatter so far measured
\citep[e.g][]{gnerucci2011}.
The origin and evolution of the scatter of the TFR is also considered as an open problem. Observational
results suggest that 
it might be driven by
systems which exhibit disturbed kinematics as a consequence of galaxy interactions
and mergers  \citep{kannappan2004, flores2006, puech2008, 
kassin2007, covington2010}.
Based on previous studies of the TFR \citep{tully1985},
\citet{weiner2006} proposed a new kinematical estimator
generating a combined velocity scale, $s_{K}$:

\begin{equation}
s_{\rm K}^{2}  = K V_{\rm rot}^{2} + {\sigma}^2,
\end{equation}

where $K$ is a constant $\le 1$ and $\sigma$ is the velocity dispersion.
The scale $s_{\rm K}$
combines rotation (associated to order motion) and pressure support 
(associated to turbulent and random motion).
\citet{kassin2007} reported that the observed sTFR
resulting from the used of $s_{0.5}$ is capable of generating
a unified relation for galaxies of all morphological types in their sample, including disturbed and merging
cases.  This TFR based in $s_{0.5}$ was also found to be remarkably tight
over $0.1 < z < 1.2$.   
By using hydrodynamical pre-prepared  merger simulations,
\citet{covington2010} showed that the kinematical indicator $s_{0.5}$ 
correlates with the total potential well of galaxies including baryons and dark matter.
A close examination of the kinematical scaling laws of galaxies led these authors to
conclude that the appropriate constant $K$ should be $\sim 0.5$.
\citet{covington2010}  also found that  the scatter of the TFR correlates with close encounters
and mergers which suggests that the  kinematics could be used to determine the merger stage
of the galactic systems \citep[see also][]{pedrosa2008}.  
Finally, these authors introduced the kinematical parameter $S= \sqrt{2} \times s_{0.5}$ which seems to be a good
tracer of the circular velocity in their simulations.
Also, \citet{weiner2006} proposed a modified version as $s_{\rm 1.0}= \sqrt{V_{\rm rot}^{2} + {\sigma}^2}$.

In the current paradigm for galaxy formation, systems assemble
hierarchically in a bottom-up fashion so that more massive galaxies
formed by the  aggregation
of smaller ones.  In this scenario, mergers and interactions between galactic systems play a major
role in the evolution of baryonic matter
within the dark matter potential wells, partially regulating the gas infall and star formation process,
which also influence the efficiency of feedback mechanisms.
In this context, cosmological N-body/hydrodynamical simulations constitute an optimum tool
to follow the evolution  of the astrophysical properties of galaxies and may help
to the understanding of the fundamental relations between them.
Our simulations have been run with a version of GADGET-3 which includes a physically-motivated Supernova (SN) feedback \citep{scannapieco2006} which is able to trigger mass-loaded gas outflows. This scheme does not require {\it ad hoc} mass
dependent parameters but naturally regulates the strength of the SN feedback in haloes of different masses \citep{scannapieco2008}. 

In \citet{derossi2010}, we studied the sTFR and bTFR in simulated galaxies with 
very well defined-dominating gaseous discs (disc-to-total gas mass ratios $D/T \ge 0.75$) formed
in a $\Lambda$-CDM universe. We found that both the sTFR and the bTFR are well reproduced by our simulations. 
Our SN feedback model works in a self-regulated way so that small systems are more largely affected than
large ones.  Our simulated sTFR is in agreement to that recently estimated by \citet{reyes2011}, in the same mass
range. While the sTFR is detected to have a bend in the slope for low stellar mass systems as a result of the SN feedback, 
the bTFR does not show the same feature except at very high redshifts. So, in this previous work, we found that the bTFR does not
store information on the impact of SN feedback for reasonable SN energies, except at very early times ($z \approx 3$).

\begin{figure}
\begin{center}
\resizebox{8.5cm}{!}{\includegraphics{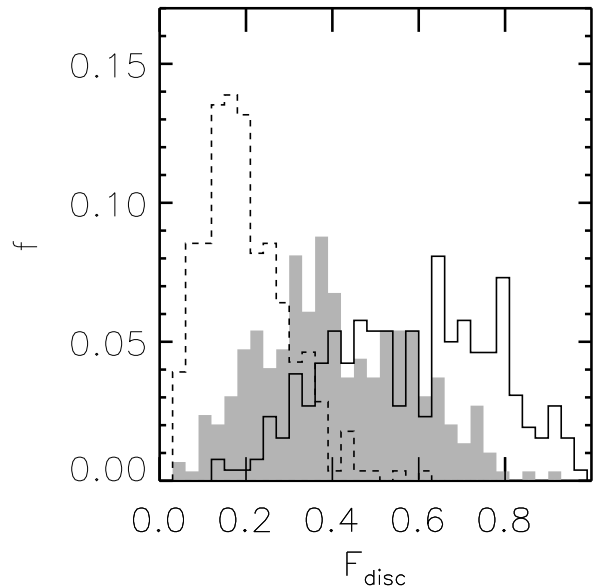}}
\end{center}
\caption[Mass fraction in the disc component at $z=0$]
{
Distributions of disc-to-total mass ratios ($F_{\rm disc}$)  estimated by using the gas (solid line), stellar (dashed line) and baryonic (shaded area) mass in simulated galaxies at $z=0$. All simulated galaxies have a surviving gaseous disc even in systems where stars are all
forming an spheroidal component.}
\label{fig:fspheroid}
\end{figure}

\begin{figure}
\begin{center}
\resizebox{8.5cm}{!}{\includegraphics{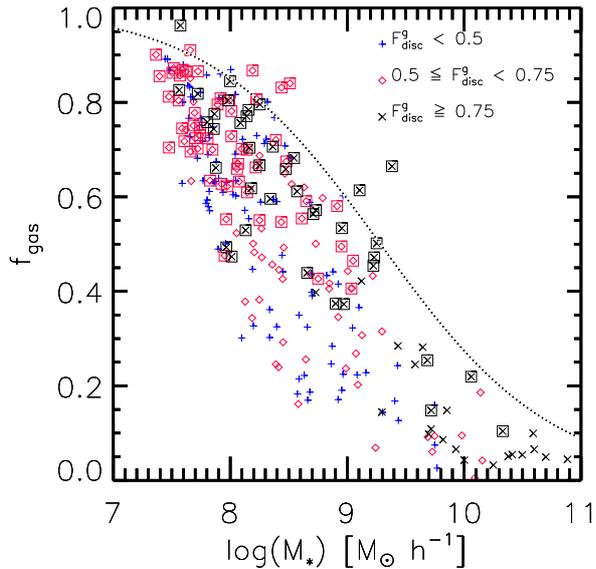}}
\end{center}
\caption[Kinematic analysis at $z=0$]
{
Gas fraction as a function of stellar mass for simulated galaxies.
Different symbols depict systems with different values of $F_{\rm disc}^{\rm g}$, as indicated in
the figure.  Symbols enclosed by squares correspond to galaxies with $F_{\rm disc}^{\rm b} \ge 0.5$.
The black dotted curve are the analytical fits to observations given
by \citet{stewart2009} at $z=0$.
}
\label{fig:pmb_vs_pm}
\end{figure}

In this paper,  we extend the work of  \citet{derossi2010} studying the 
 gas kinematics of all galaxies with surviving gaseous discs, even those systems dominated by spheroidal components, 
using the same set of 
cosmological hydrodynamical simulations.
Particular attention is paid to the study of the origin and evolution of the scatter in the TFR 
considering  the impact of mergers and interactions as well as the importance of gas inflows and outflows.
The rotation curves are derived from the analysis of the surviving gaseous discs with the aim 
at determining the best gas-phase kinematical  tracer for the gravitational potential. 
We assumed that, even if the surviving gaseous disc is tenuous, it will trace
the potential well in which it inhabits.

In section \ref{sec:simus}, we introduce the numerical simulations studied in this
work, the galaxy catalogue and the methods developed to perform the analysis of gas
kinematics.  In section \ref{sec:localTFR}, we present
a discussion of the local TFR and various kinematic gas indicators for galaxies
of different morphologies. 
In section \ref{sec:evolution}, we investigate the fingerprints of the hierarchical aggregation
of the structure on the TFR-plane.  The conclusions of the work
are summarised in section \ref{sec:conclusions}.

\section{Numerical Experiments}
\label{sec:simus}

The simulation analysed in this work  was performed by using the chemical code {\small GADGET-3},
an update of {\small GADGET-2} optimised for massive parallel simulations of highly inhomogeneous
systems \citep{springel2003, springel2005}.
Our version of the code includes treatments for metal-dependent radiative cooling, 
stochastic star formation,  a multiphase model for the interstellar medium (ISM) and
a  SN feedback scheme \citep[][]{scannapieco2005, scannapieco2006}.  
The multiphase model allows the coexistence of diffuse and dense gas phases,
improving  the description
of the ISM  \citep{scannapieco2006}. 
The SN feedback model is able to trigger mass-loaded galactic
outflows without introducing mass-dependent parameters.

\begin{figure*}
\begin{center}
\resizebox{6.5cm}{!}{\includegraphics{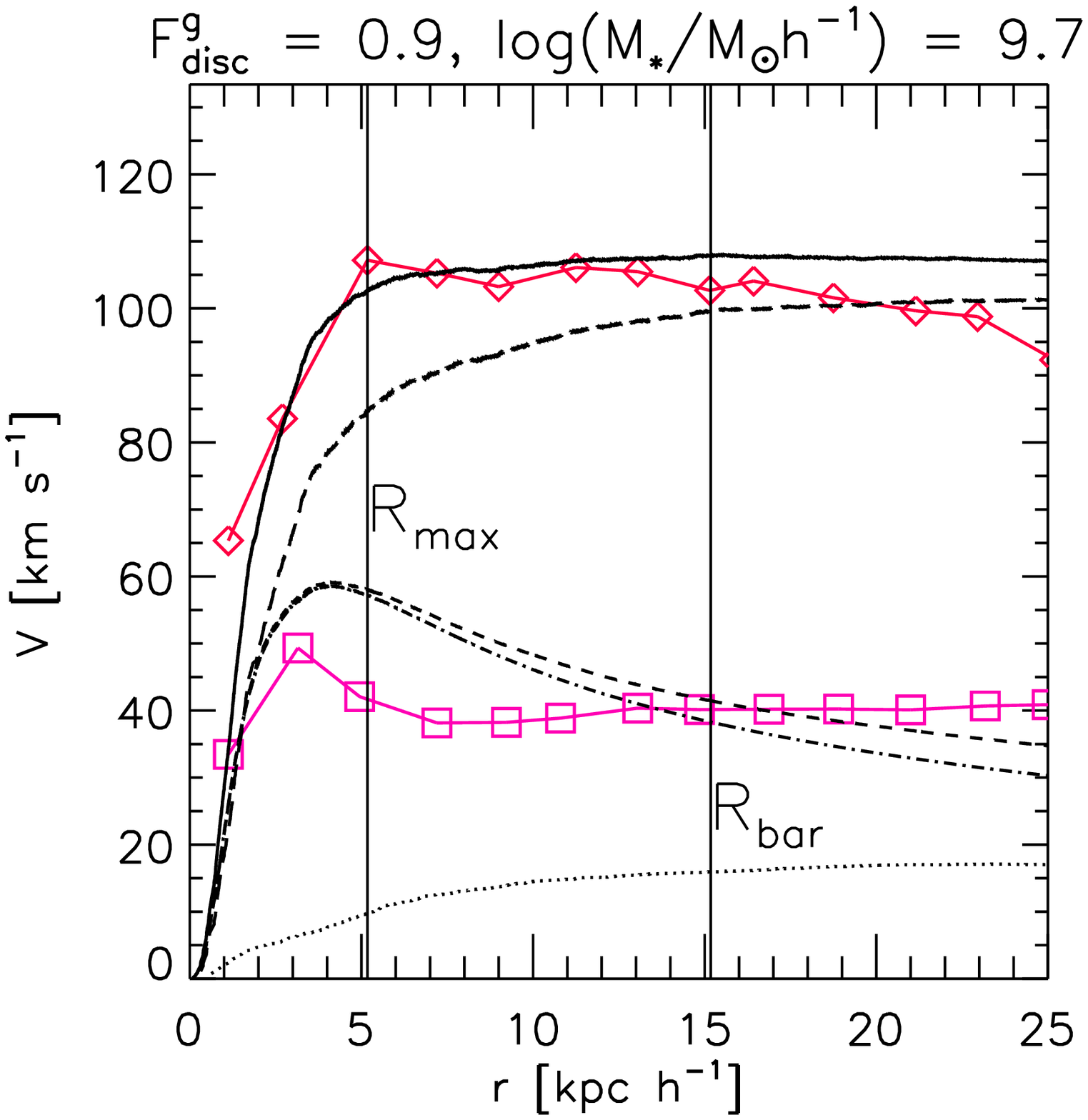}}
\resizebox{6.5cm}{!}{\includegraphics{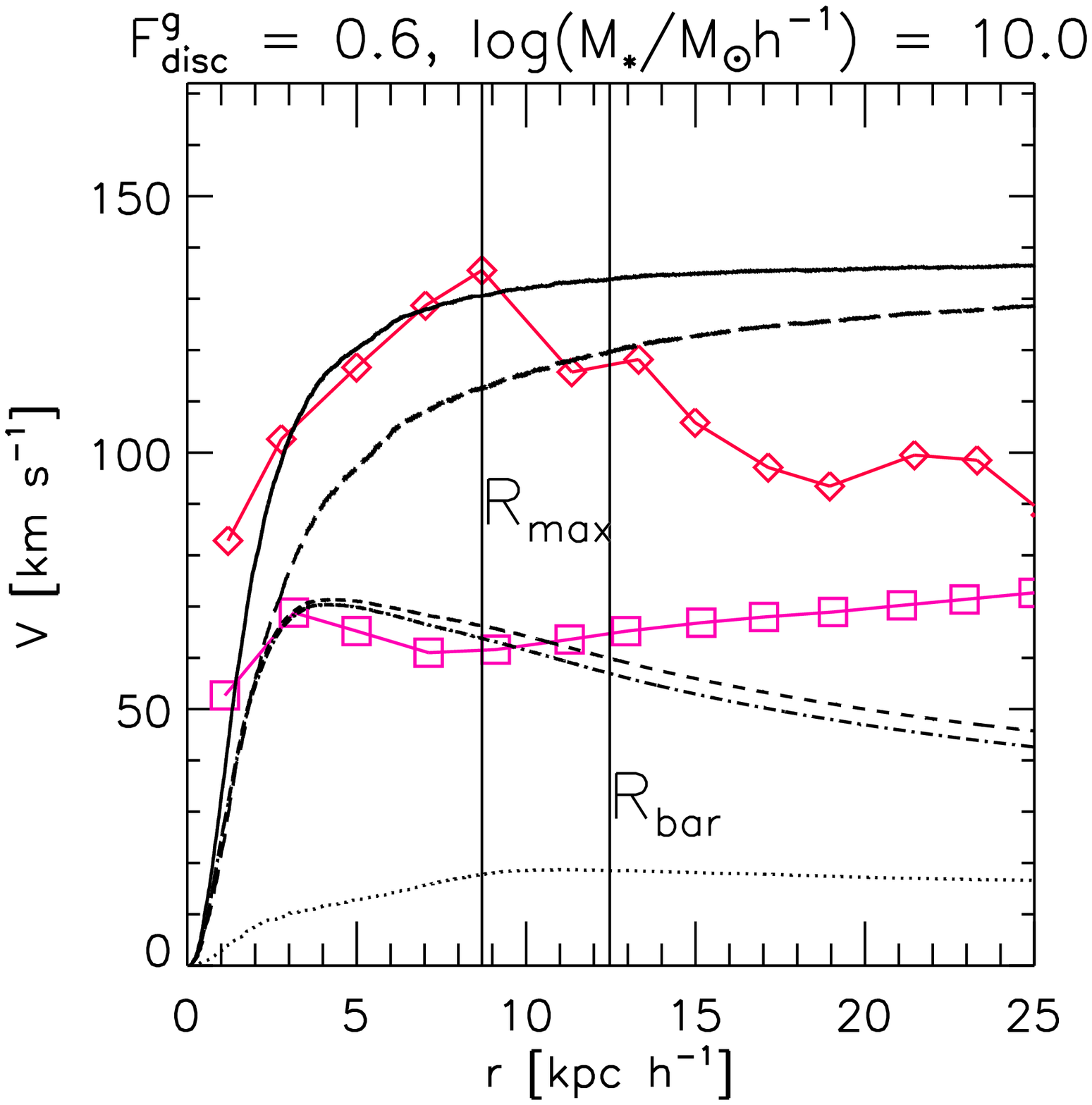}}\\
\resizebox{6.5cm}{!}{\includegraphics{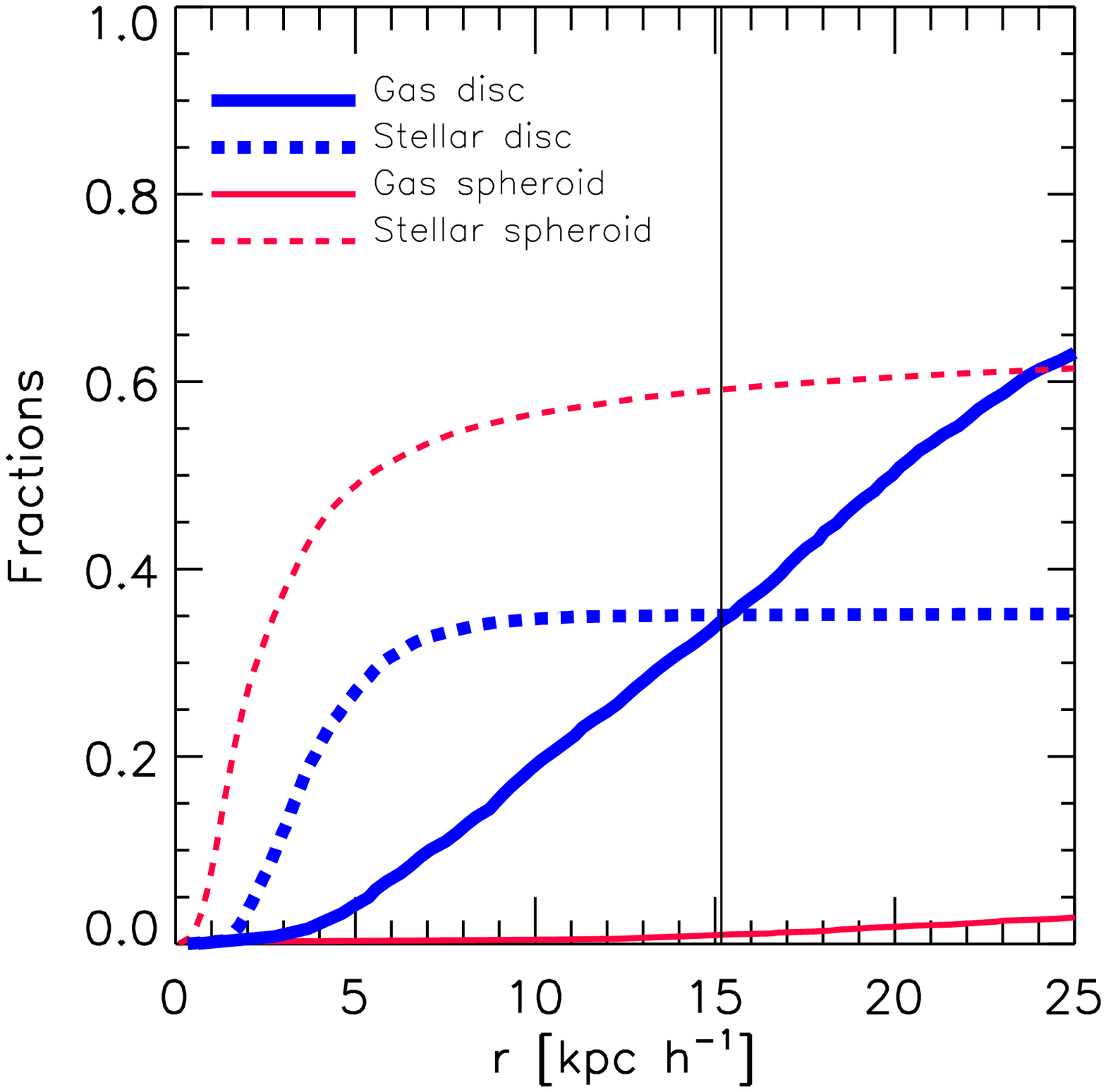}}
\resizebox{6.5cm}{!}{\includegraphics{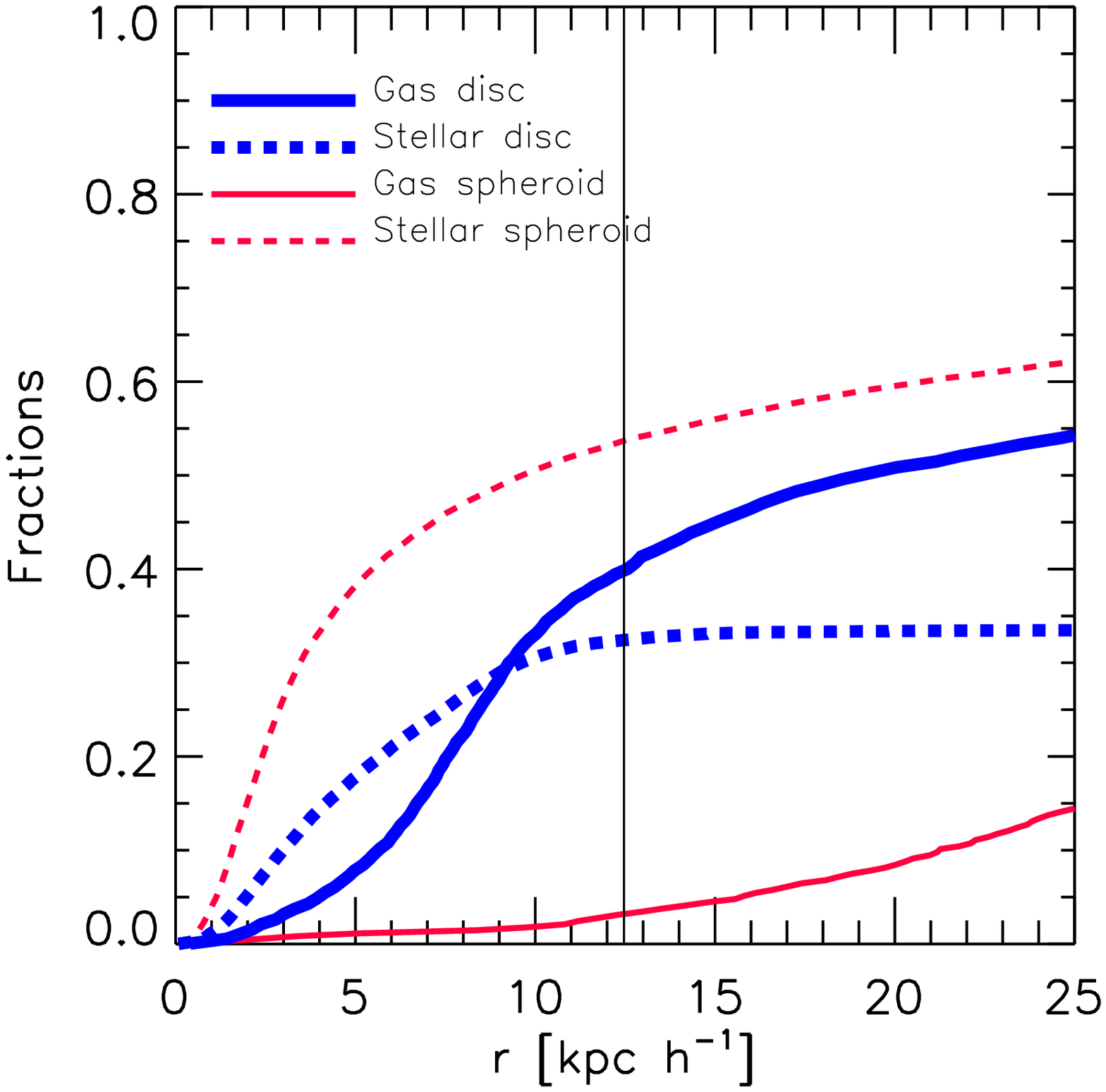}}
\end{center}
\caption[Rotation curve and morphology]
{
Upper panels: Rotation curves for two typical 
simulated galaxies G1 and G2 (left and right panels, respectively) at $z=0$ with different fractions of gas supported by rotation 
  ($F_{\rm disc}^{\rm g}$).  The total circular velocity is shown 
with solid line.
The circular velocities contributed by 
the gas, stellar, baryonic and dark components are represented with
dotted, dot-dashed, dashed and long-dashed lines, respectively.  
The mean rotation curve  (diamonds)
and the mean dispersion velocity curve (squares) for the gas-phase are also plotted.   
The vertical lines indicate the baryonic radius ($R_{\rm bar}$)
and the radius where the mean rotation velocity reaches a maximum value ($R_{\rm max}$).
Lower panels: Gas (solid lines) and stellar (dashed lines) mass in the
spheroidal (red) and disc (blue) components as a function of radius, normalised to the corresponding 
total gas or stellar mass
within  $3 R_{\rm bar}$.  The vertical line depicts $R_{\rm bar}$.
}
\label{fig:rotcurv}
\end{figure*}

The chemical evolution prescription used in this code was developed by \citet{mosconi2001} 
and, later on, adapted by \citet{scannapieco2005} for
{\small GADGET-2}. The model describes the  enrichment by
Type II (SNII) and Type Ia (SNIa)  Supernovae 
according to the chemical yield prescriptions of \citet{woosley1995}
and \citet{thielemann1993}, respectively.
We adopted a standard Salpeter Initial Mass Function with a lower and upper
mass cut-offs of 0.1 $\rm M_{\odot}$ and 40 $\rm M_{\odot}$, respectively.
It is assumed that each SN event generates $0.7 \times 10^{51}$ erg.
For SNIa, we assumed a time-delay for the ejection of material randomly chosen within
$[0.1,1]$ Gyr. We assumed that SNII evolved within the same time interval in which  they are
formed.
Metals are distributed within the neighbouring gas particles weighted by the
smoothing kernel as proposed by \citet{mosconi2001}.

We  use numerical hydrodynamical simulations consistent with  a
$\Lambda$-CDM universe with
$\Omega =0.3, \Lambda =0.7, \Omega_{b} =0.04$, a normalisation of
the power spectrum of ${\sigma}_{8} = 0.9$ and
$H_{0} =100 \, h$ km s$^{-1} {\rm Mpc}^{-1}$ with  $h=0.7$.
Although this set of parameters differs slightly from WMAP-7, the variations have no
important consequences on our results.
We simulated a typical field region of the Universe in a comoving 
cubic volume of 10 Mpc $h^{-1}$ side length.
The simulation studied in this paper was run with $2 \times 230^3$  particles
obtaining an initial mass  of
$9.1 \times 10^5 \ {\rm M_{\odot}} h^{-1}$ for gas particles,
and a mass of $5.9 \times 10^6 \ {\rm M_{\odot}} h^{-1}$ for the dark matter particles.
It is the so-called S230 in  \citet{derossi2010}.

In \citet{derossi2010}, we showed that the dynamical properties
of the galaxies in this sample are robust against numerical resolution and small changes in
the parameters of the SN-feedback model.
We also found that this simulation predicts correlations between
the dynamical properties of galaxies in general good agreement
with observations, providing a suitable sample  to study the origin and evolution
of the TFR in a cosmological context.
In particular, we found that the stellar mass fractions
of galaxies residing within 
dark matter haloes are in good agreement with estimations derived from  semi-empirical models 
\footnote{ Note that those semi-empirical studies used different Initial Mass
Functions (Chabrier/Kroupa) to \citet[][Salpeter]{derossi2010}
and, hence, for the same luminosity implies some 40\% less mass locked in stars. 
Even taking into account these differences,
the simulated stellar mass fractions remain within the observational predicted range.} 
based on
the observed luminosity function \citep[][]{guo2010, moster2010}.
In \citet{derossi2010}, it was also proven that the adopted SN feedback model is able to reproduce successfully 
the observed bend of the sTFR at $\sim 100 \, {\rm km} \, {\rm s}^{-1}$ 
\citep[e.g.][]{mcgaugh2000, amorin2009, torres2011} as a consequence
of its high efficiency at regulating the star formation process of galaxies in haloes with shallow potential wells
\citep{derossi2010}. On the contrary, the bTFR shows no SN feedback imprinted feature at  low redshift.

\subsection{The galaxy catalogue}
\label{sec:sample}

The virialized structures in the simulated box are identified
by a standard friends-of-friends technique
and the substructures within each dark matter halo
are then individualised 
by using the SUBFIND algorithm \citep{springel2001}.
In order to diminish numerical problems, we only  analysed substructures
 with a total number of particles $N_{\rm sub} > 2000$ from  $z \sim 4$.
This gives a total number of galaxies which ranges from
150 at $z=3$ to 309 at $z=0$.

The dynamical properties of the simulated galaxies have been estimated within the baryonic
radius $R_{\rm bar}$, defined as the one which encloses 83 per cent of
the baryonic mass of the system.
In these simulations, a typical Milky-Way type galaxy is resolved with around
$10^5$ total particles within $R_{\rm bar}$.
It is also worth noting that for our simulated galaxies more than 90\% of the gas inside $R_{\rm bar}$
have $T < T_c = 8\times 10^4$ K, with  $T_c$ being the critical temperature used by
the multi-phase model of the ISM \citep{scannapieco2006}. 
Therefore, for the sake of simplicity,
we did not distinguish between a cold and hot component in the analysis 
of gas dynamics as the mean behaviour is almost entirely determined by the cold gas.

\begin{figure*}
\begin{center}
\resizebox{19.cm}{!}{\includegraphics{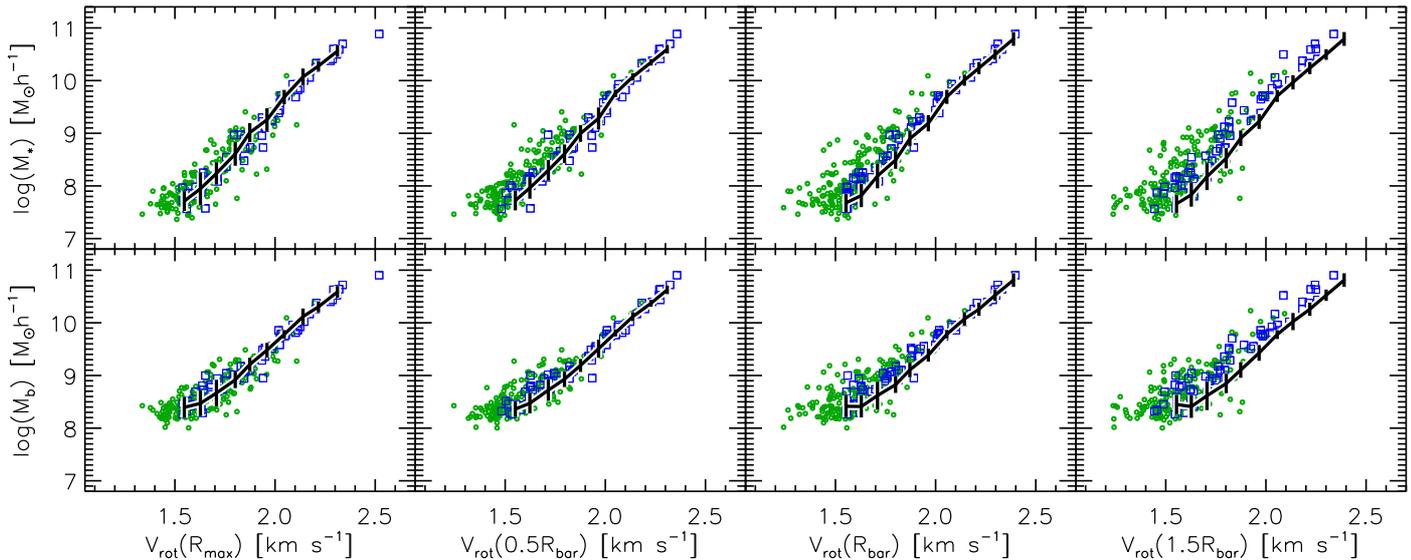}}
\end{center}
\caption[TFR at $z=0$]
{
Simulated sTFR (upper panels) and bTFR (lower panels) at $z=0$ by using $V_{\rm rot}$  estimated
at different radii
as indicated in the labels.
The green circles depict systems with $F_{\rm disc}^{\rm g} < 0.75$, while the blue squares
represent systems with $F_{\rm disc}^{\rm g} \ge 0.75$.  
The black solid lines indicate
the mean relations estimated by using $V_{\rm circ}$ measured at the corresponding radii and their 
standard deviations. 
}
\label{fig:TFR}
\end{figure*}

\subsection{Kinematic analysis of simulated galaxies}
\label{sec:kinematics}

For a given simulated galaxy, 
we calculated its circular velocity as 
$V_{\rm circ} = \sqrt{G M(r) / r}$,  
where $M(r)$ is the total mass enclosed within the radius, $r$.  
In equilibrium, for baryons dominated by rotation,
the rotation velocity $V_{\rm rot} \sim V_{\rm circ}$, which
is a measure of the potential well of the system.
For each gas (stellar) particle $i$ within a given galaxy,
we defined the rotation velocity 
($V_{{\rm rot},i}$) as its tangential velocity on the plane
perpendicular to the 
total gas-phase (stellar-phase) angular momentum ${\vec{J}}_{\rm g}$ (${\vec{J}}_{\rm s}$) of the system. 
We estimated the averaged values of 
the velocity components
for particles at equally-spaced radial bins in the gas and 
in the stellar components.
Then, we calculated the dispersion velocity (${\sigma}_i$) for each particle $i$ inside 
these bins.
Finally, following \citet{scannapieco2005}, we used the ratio between
the dispersion and the rotation velocity of each baryonic particle (${\sigma}_i / V_{{\rm rot},i}$)
to define the disc and spheroid component of the given galaxy.  
If ${\sigma}_i / V_{{\rm rot},i}  < 1$, 
we considered that the particle $i$ belongs to the disc component, otherwise it was assumed that it
belongs to the spheroid component
\footnote{We checked that this criterion produces similar results to those
obtained by applying  the methods used by \citet{abadi2003} or \citet{scan09} or \citet{tiss2011}.}.  
Note that our definition implies that the spheroid component is made of all particles
dominated by dispersion  
so that it includes both the bulge and the inner stellar halo.

In order to characterise the morphology of simulated galaxies, we estimated $F_{\rm disc}^{\rm g}$ ($F_{\rm disc}^*$) as
the mass fraction of the total gas (stellar) mass within $3 R_{\rm bar}$ which is supported by rotation 
(i.e. the $D/T$ ratio for the gas or stellar phase, respectively).
We found that, in general, our sample of simulated galaxies is constituted
by systems with dominating stellar spheroids and thick stellar discs while 
the gas mass fractions associated to the spheroid or the disc are very assorted.
This behaviour can be appreciated  more clearly in Fig. \ref{fig:fspheroid}, where we show the histograms 
of $F_{\rm disc}^{\rm g}$ (solid line), $F_{\rm disc}^*$ (dashed line) and a similarly defined fraction for the baryonic mass 
($F_{\rm disc}^{\rm b}$, shaded area) for
simulated galaxies at $z=0$
\footnote{$F_{\rm disc}^{\rm g}$, $F_{\rm disc}^*$ and $F_{\rm disc}^{\rm b}$ are equivalent to the commonly
used $D/T$ ratio for the gas, stellar and baryonic components.}.
We can see that the percentage of the stellar mass dominated by rotation does not exceed
60\% for any of the analysed galaxies
while, in  the case of the  gas component,
we obtained percentages ranging 20\% to $\sim 100\% $.  When considering the baryonic phase as a whole,
the percentage of mass dominated by rotation exhibits an intermediate behaviour with
percentages ranging between $10\%$ and $80\%$.
 Most of the simulated galaxies in our sample have an important spheroidal component and all of them have a surviving 
gaseous discs from which it is possible to estimate rotation curves. Our goal here
is to analyse if, regardless of 
galaxy morphology,
the gaseous discs can defined a TFR consistent with observations and
if the corresponding rotation curves are able to trace the potential wells of their galaxies.

\begin{figure}
\begin{center}
\resizebox{7.5cm}{!}{\includegraphics{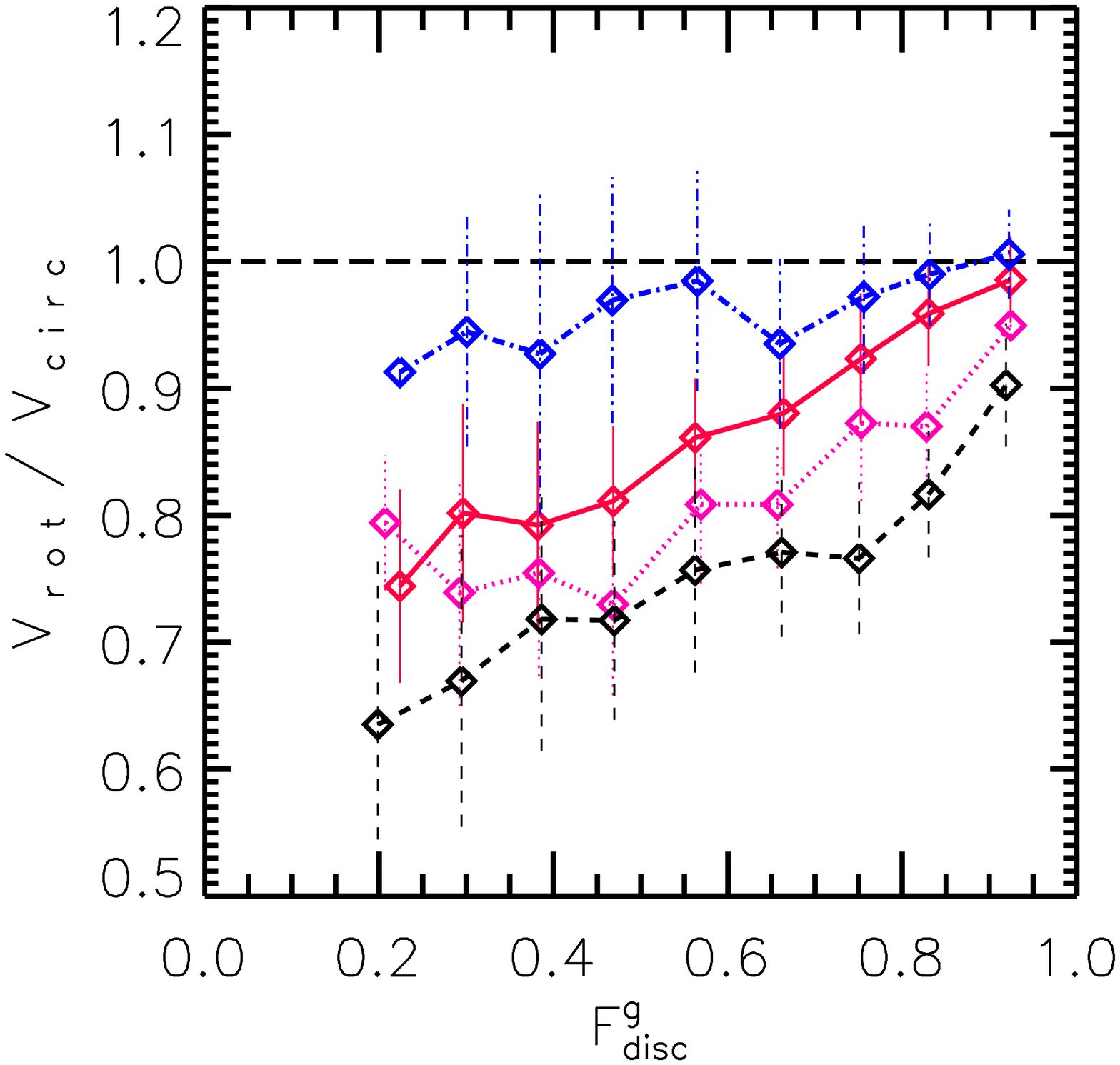}}\\
\resizebox{7.5cm}{!}{\includegraphics{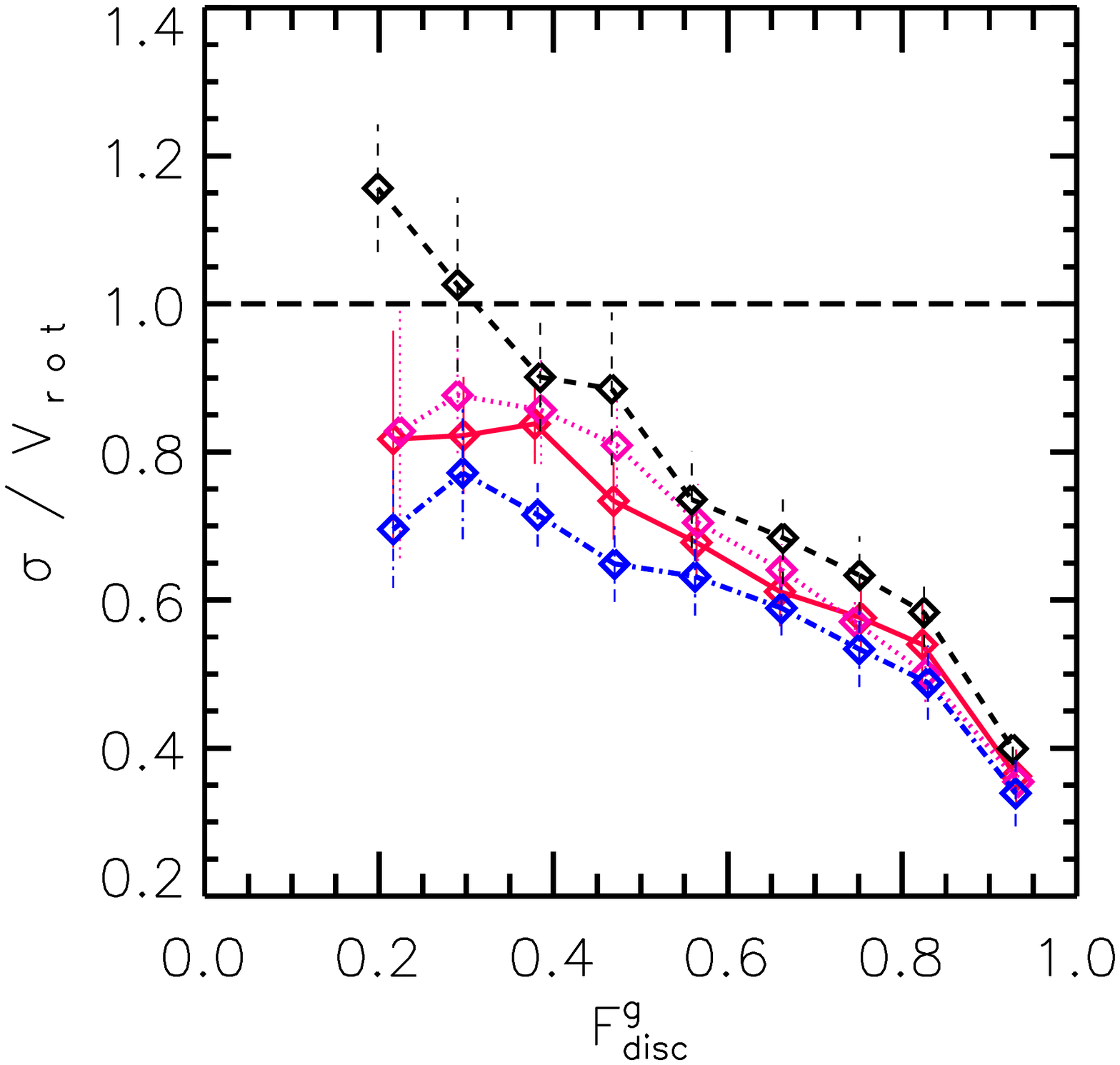}}
\end{center}
\caption[Kinematic analysis at $z=0$]
{
$V_{\rm rot} / V_{\rm circ}$ and $\sigma / V_{\rm rot}$  as a function of the fraction of the gas mass supported
by rotation and at different radii in the simulations at $z=0$: 
$R_{\rm max}$ (dot-dashed line), $0.5 R_{\rm bar}$ (solid line), $ R_{\rm bar}$ (dotted line)  and $1.5 R_{\rm bar}$ (dashed line).  
Error bars correspond to the standard deviations.
It is interesting to note that $V_{\rm rot}$ at $R_{\rm max}$ can approximate
$V_{\rm circ}$ with good accuracy ($> 90 \%$) for all simulated systems.  
}
\label{fig:VrVc_vs_pm}
\end{figure}

As shown in Fig. \ref{fig:pmb_vs_pm},  the  $f_{\rm gas}$ of our simulated galaxies decreases with stellar mass, consistently with observations, although, at a given stellar mass, 
they have lower gas fractions than observed ones \citep[e.g.][]{stewart2009, reyes2011} as a consequence
of the very efficient SN feedback assumed in this model \citep[e.g.][]{derossi2010}.
We also appreciate that
simulated galaxies with dominant baryonic discs ($F_{\rm disc}^{\rm b} \ge 0.5$) are biased to lower stellar masses and 
high gas fractions indicating that the baryonic discs are formed by significant gas masses.
In this figure, we also highlighted the simulated galaxies according to the parameter $F_{\rm disc}^{\rm g}$ 
($F_{\rm disc}^{\rm g} < 0.5$, $0.5 \le F_{\rm disc}^{\rm g} < 0.75$, $F_{\rm disc}^{\rm g} \ge 0.5$) to show
the large variety of  gas disc-type galaxies coming out of the simulations. While there are not massive stellar galaxies
($M_* \ga 10^{10} M_\odot h^{-1}$) with important stellar
disc components, the left-over gas mass  is mainly dominated by rotational motions.  We have no system dominated
by stellar discs as indicated by the lack of population with $F_{\rm disc}^{\rm g} < 0.5$ and $F_{\rm disc}^{\rm b} \ge 0.5$.

Normal spiral galaxies have their gaseous and stellar disc components aligned. Cases of misalignment are generally considered the product of interactions or recent mergers which perturbed the kinematics of the systems \citep{snaith2012}.  
Therefore, in order to have a more clear picture of the kinematics of baryons in our sample,
we calculated the cosines of the angles ($\alpha$) determined by the angular momenta of the gas discs (${\vec{J}}_{\rm g}$) and
those of the stellar ones (${\vec{J}}_{\rm *}$).
In the case of galaxies with dominant baryonic discs ($F_{\rm disc}^{\rm b} \ge 0.5$), we estimated
these cosines taken into account only the mass associated to the disc component while, for systems
with dominant baryonic spheroids ($F_{\rm disc}^{\rm b} < 0.5$),
we considered the whole baryonic mass.
Our results show that 
$\cos ( \alpha ) > 0.95$ for all systems with $F_{\rm disc}^{\rm b} \ge 0.5$, 
showing that gaseous and stellar discs are very well aligned, consistently with disc-like galaxies.  
In the case of galaxies with $F_{\rm disc}^{\rm b} < 0.5$, 
${\vec{J}}_{\rm g}$ and ${\vec{J}}_{\rm *}$ are not always aligned, forming angles larger than $\pi / 4$ in some cases,
which reveals different levels of disturbance in the baryonic kinematics.
For future discussion, we will define three subsamples:
S1, constituted by systems with dominant baryonic discs,
which are more similar to real spiral galaxies; 
S2, formed by systems with dominant baryonic spheroids and aligned gas discs 
($F_{\rm disc}^{\rm b} < 0.5$ and $\cos ( \alpha ) > 0.7$), and 
S3, including galaxies also with  dominant baryonic spheroids but misaligned gas discs
($F_{\rm disc}^{\rm b} < 0.5$ and $\cos ( \alpha ) \le 0.7$).
The percentages of galaxies corresponding to S1, S2 and S3 are 33\%, 31\% and 36\%, respectively.

As  the main goal of this work is the analysis of the gas kinematics in the disc component, we will characterise
simulated systems by using only $F_{\rm disc}^{\rm g}$ but, when necessary, we will also refer to the subsamples
S1, S2 and S3 in order to get a more complete picture.
Note that in \citet{derossi2010}, we restricted our study to galaxies with 
$F_{\rm disc}^{\rm g} \ge 0.75$ so that the gaseous disc components were important with well-behaved  rotation curves. 
As mentioned before, in this work, we extend that analysis to the
whole sample of simulated galaxies with a surviving gas disc.

\subsubsection{Rotation curves}

In the upper panels of Fig. \ref{fig:rotcurv}, we show the rotation  and
circular velocities for two typical simulated galaxies at $z=0$. 
The left panels exhibit the results for a galaxy with a stellar mass of 
$M_* = 10^{9.7}  M_{\odot} h^{-1}$ and $F_{\rm disc}^{\rm g} = 0.9$ (G1), while in the
right panels we display those  of a galaxy with a stellar mass of
$M_*  = 10^{10}  M_{\odot} h^{-1}$ and $F_{\rm disc}^{\rm g} = 0.6$ (G2).
G1 belongs to the subsample S1 
(i.e. systems more similar to real spiral galaxies) 
and G2 to the subsample S2
(i.e. systems with dominant baryonic spheroids, with aligned gas discs).
As can be seen, G1 has  a more important gas-phase disc component with a 
rotation curve tracing the circular velocity remarkably well until $R_{\rm bar}$,
while in the case of G2, it drops around $R_{\rm max}$, departing significantly from the circular velocity.  
Note that, in both cases, the decrease in the rotation
curve is accompanied by an increase of the gas mean dispersion velocity, as expected.

In the lower panels of Fig. \ref{fig:rotcurv}, we can appreciate
the gas (solid lines) and stellar (dashed lines) mass fractions associated to the
spheroidal (red) and disc (blue) components  as a function of radius for G1 and G2.
The gas (stellar) mass in each component is normalised to the total gas (stellar) mass
enclosed by $3 R_{\rm bar}$.  
We see that,  in the case of G1, the stars are  more concentrated than in G2.

\begin{figure*}
\begin{center}
\resizebox{8.5cm}{!}{\includegraphics{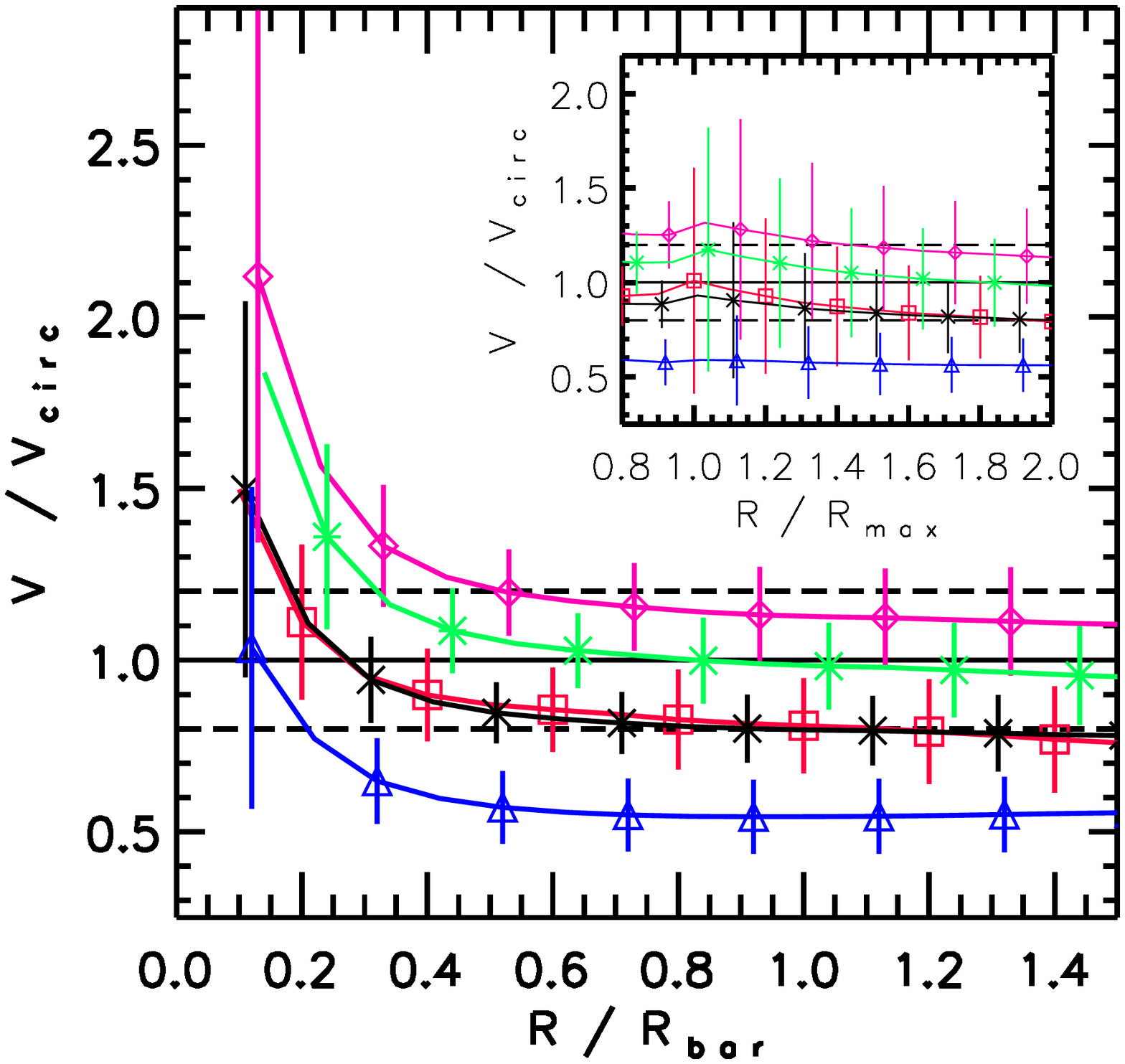}}
\resizebox{8.5cm}{!}{\includegraphics{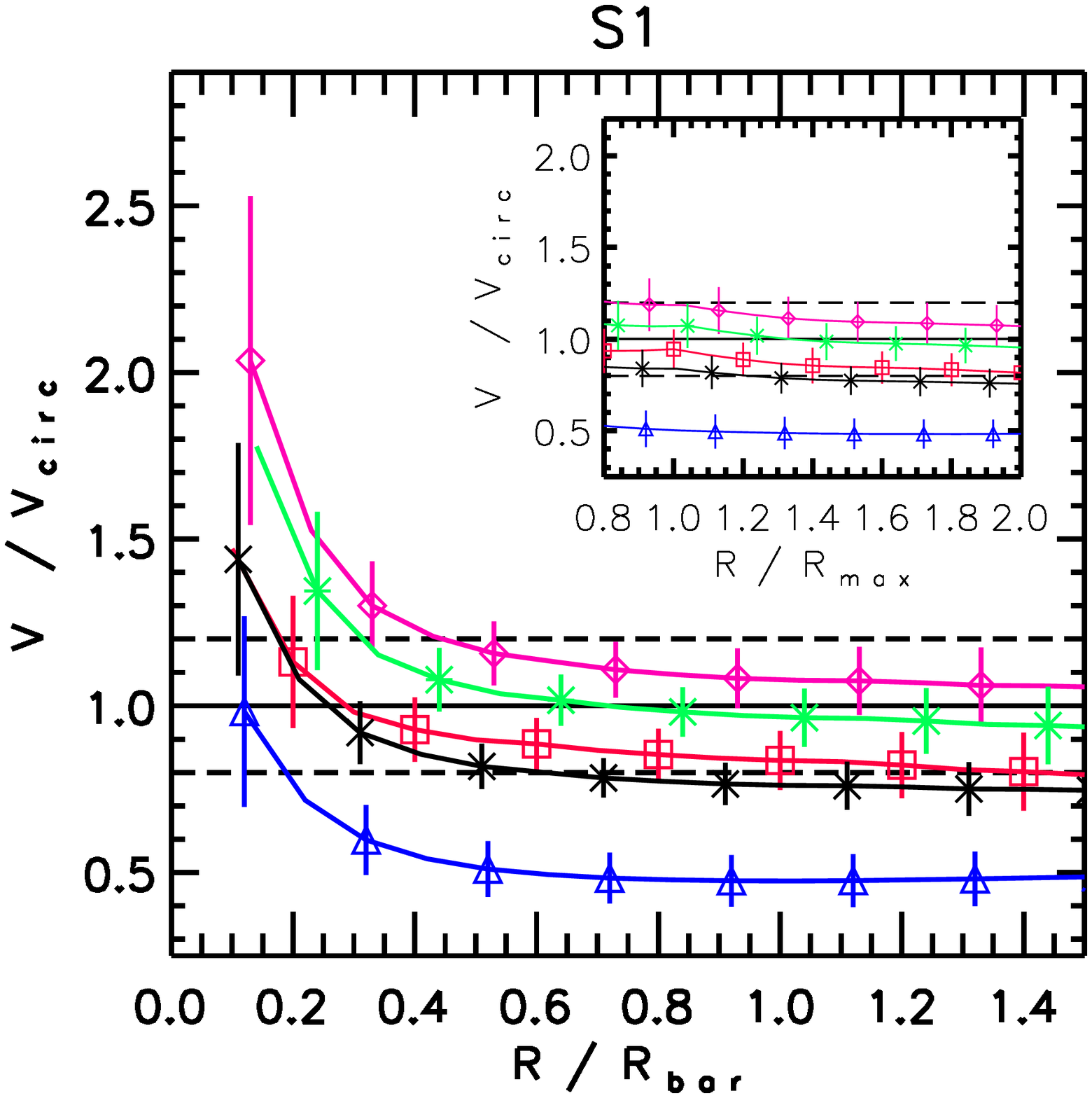}}\\
\resizebox{8.5cm}{!}{\includegraphics{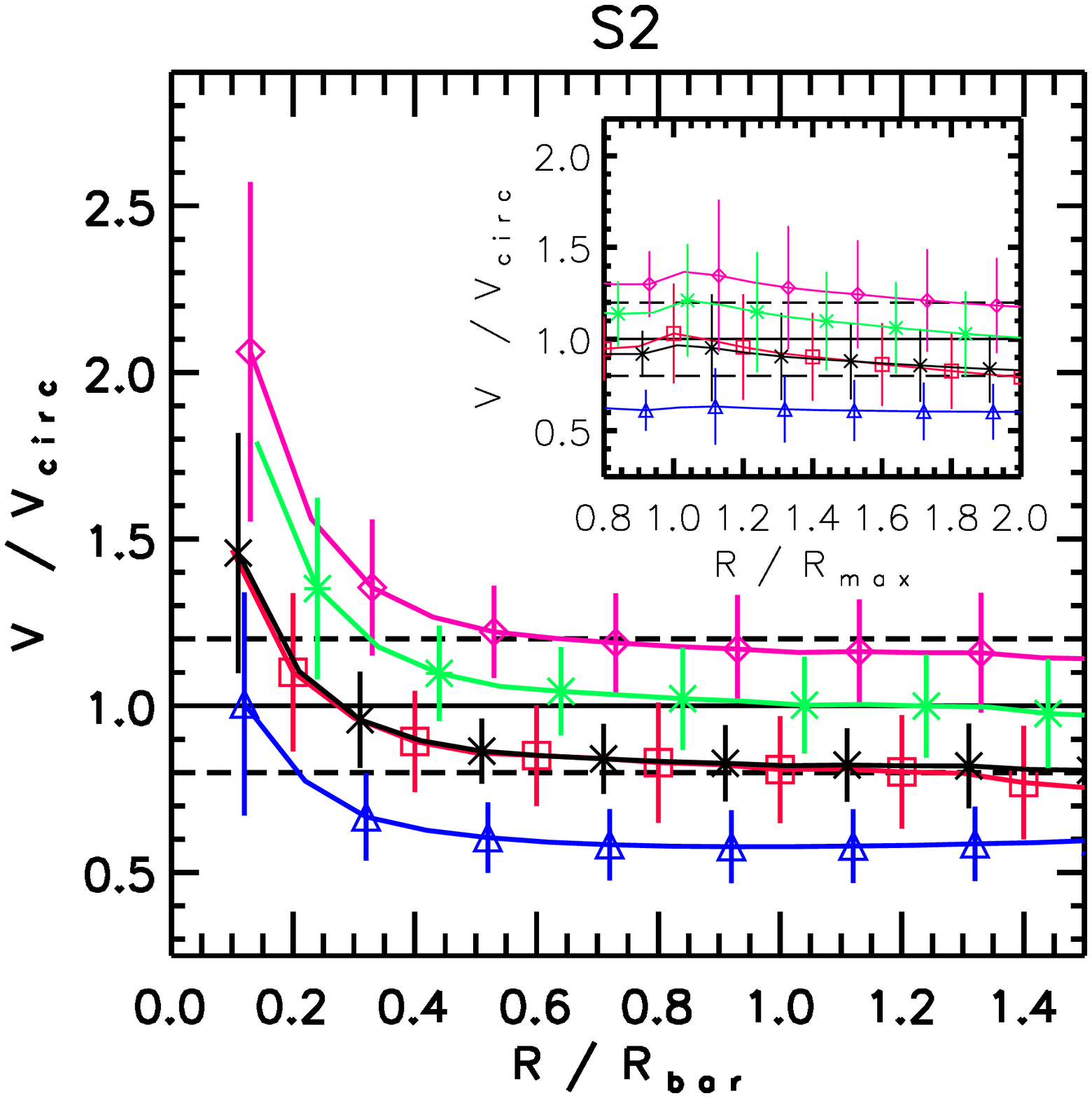}}
\resizebox{8.5cm}{!}{\includegraphics{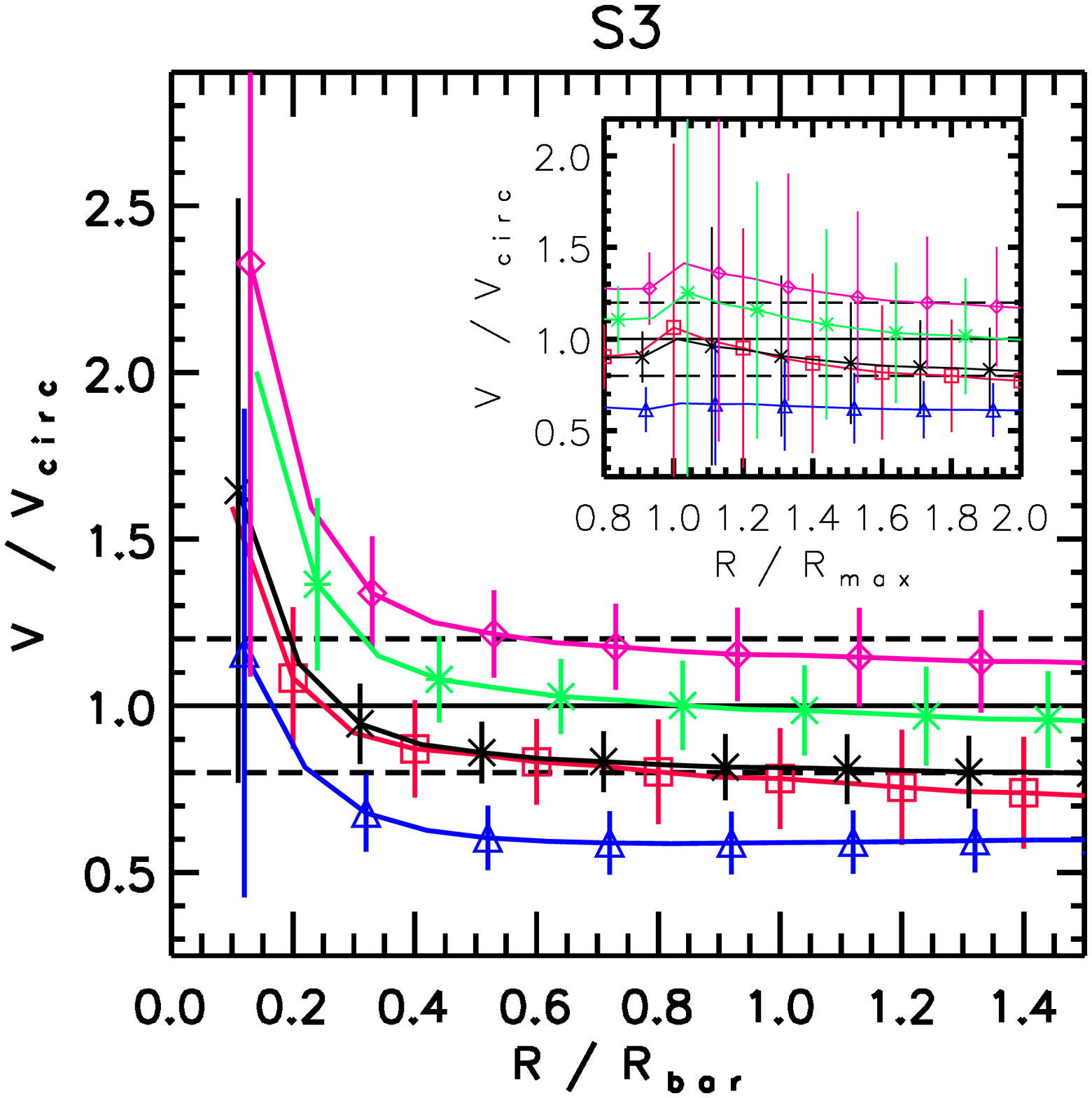}}
\end{center}
\caption[$V_{\rm i} / V_{\rm circ}$ at $z=0$]
{
$V_{\rm i} / V_{\rm circ}$ at $z=0$ as a function of radius in the simulations
for $V_{\rm i}$: $\sigma$ (blue triangles), $V_{\rm rot}$ (red squares), $s_{\rm 0.5}$ (black crosses),
$s_{\rm 1.0}$ (green asterisks) and $S$ (pink diamonds).  
Each symbol represents the averaged value over the entire sample of galaxies  
(upper left panel) and the subsamples S1, S2 and S3 defined in section \ref{sec:kinematics}
(upper right panel, lower left panel and lower right panel, respectively).
The corresponding standard deviations are also shown.  Note that the radii are given normalised to
$R_{\rm bar}$ in the main panels while, in the insets, they are given normalised to $R_{\rm max}$.
The horizontal solid line indicates where
$V_{\rm i}$ agrees with $V_{\rm circ}$, while the dashed horizontal lines denote
the range where $V_{\rm i}$ does not depart more than 20\% from $V_{\rm circ}$.
}
\label{fig:ViVc_vs_r}
\end{figure*}

\section{The Tully-Fisher relation}
\label{sec:localTFR}

The kinematical analysis performed for G1 and G2 in section \ref{sec:kinematics}
was extended to our whole sample
of simulated galactic systems at $z=0$.
We constructed the sTFR and bTFR by using the rotation velocity and the circular velocity
estimated  at different radii: $R_{\rm max}$,  $0.5 R_{\rm bar}$, $R_{\rm bar}$ and $1.5 R_{\rm bar}$,
as shown  in  Fig. \ref{fig:TFR}.
Our simulated TFR agrees very well with that reported by \citet{reyes2011} over the whole observed mass  range 
($\approx 10^9 - 10^{11} M_{\odot} h^{-1}$).
Our SN feedback model is able to reproduce the observed sTFR and bTFR
even for these large stellar masses without the need to resort to other physical processes 
\citep[see][for different results]{mccarthy2012}.
As shown in \citet{derossi2010}, this SN feedback model can also predict the observed
bend of the sTFR at smaller stellar masses $M_* < 10^{9} M_{\odot} h^{-1}$ 
\citep[e.g.][]{mcgaugh2000, amorin2009}.
Indeed, the good agreement of our simulated TFRs with observations of large and small stellar masses is the result of the action
of the self-regulated SN feedback adopted in our models, which is able to regulate the star formation process in systems of different potential wells without
the need to introduce {\it ad hoc} mass-dependent parameters.

In the case of the gaseous disc-dominated systems (blue squares in Fig. \ref{fig:TFR}, $F_{\rm disc}^{\rm g} \ge 0.75$, 23\% of the total sample), the good agreement between the TFR obtained
by using $V_{\rm rot}$  and that estimated by using  $V_{\rm circ}$ at different radii, shows that the 
former is a good measure of the total potential well.
For these systems, we obtained that the scatter in the sTFR and bTFR decreases 
when $V_{\rm rot}$ is estimated at $\sim 0.5 R_{\rm bar}$, in agreement with observations \citep{yegorova2007}.
Also, for these systems, we found that $V_{\rm rot}(R_{\rm max}) \approx V_{\rm rot}(R_{\rm bar})$
consistently with the findings of \citet{persic1995}.

In the case of systems with more important 
gas spheroidal components (green circles in Fig. \ref{fig:TFR}, $F_{\rm disc}^{\rm g} < 0.75$), their rotation velocities 
tend to be smaller than the  $V_{\rm circ}$ at a given mass. This is accompanied by
 the increase in  gas velocity dispersion. 
To quantify the behaviour of the kinematical estimators, in  Fig. \ref{fig:VrVc_vs_pm}
we show   $V_{\rm rot} / V_{\rm circ}$  and  $\sigma / V_{\rm rot}$ 
as a function of $F_{\rm disc}^{\rm g}$, measured at   different radii ($\sigma$ is estimated by considering all gas
particles within a given radius).
When the disc dominates the gas-phase component,
$V_{\rm rot}$ can approximate $V_{\rm circ}$ with good accuracy (larger than $\sim 90\%$).
However, as $F_{\rm disc}^{\rm g}$ decreases, $V_{\rm rot} / V_{\rm circ}$ becomes smaller
and $\sigma / V_{\rm rot}$ gets larger as the gas component is more dominated by dispersion, showing more important variations when measured  at larger radii. 
For galaxies with gas component dominated by dispersion ($F_{\rm disc}^{\rm g} < 0.4$), we found 
a mean value of  $\sigma / V_{\rm rot} \sim 0.7$ measured at $R_{\rm max}$, 
in agreement with recent observational results by \citet{catinella2011}.

We recalculated these trends for  the subsamples S1, S2 and S3
(see section \ref{sec:kinematics}),  obtaining similar global results. In the case of  $\sigma / V_{\rm rot}$ , regardless of the
alignment between stellar and gaseous angular momenta,
galaxies with dominating baryonic spheroids have also more disturbed gas dynamics, implying higher ratios than
those for systems with baryonic dominating disc,  at
a given $F_{\rm disc}^{\rm g}$.

As we note from Fig. \ref{fig:VrVc_vs_pm}, 
a very interesting result from our model is that $V_{\rm rot}$ at $R_{\rm max}$ seems to be the best proxy
for the circular velocity as it shows no significant dependence on
$F_{\rm disc}^{\rm g}$:
$V_{\rm rot}$ is always within 90\% of $V_{\rm circ}$ for all the systems. 
This trend is even better
for gas systems dominated by rotation. 
We checked these trends for S1, S2 and S3, finding similar results.
Regardless of the relative
angular momentum orientation between stellar and gas components, $V_{\rm rot} $ at $R_{\rm max}$
provides, on average, the best indicator for the total circular velocity, although 
with high standard deviations 
for systems with $F_{\rm disc}^{\rm b} \lesssim 0.5$.

As discussed in the Introduction, \citet{kassin2007} and  \citet{covington2010} 
proposed
new kinematical estimators based on observational and numerical results, respectively,
   to account for the disordered motions in the gas component with
the aim at explaining the origin of the scatter in the TFR. 
We tested these estimators  in our cosmological simulations. 
In Fig. \ref{fig:ViVc_vs_r}, we can appreciate
$V_{\rm i} / V_{\rm circ}$ at $z=0$ as a function of radius
where  $V_{\rm i}$ could be either  $\sigma$, $V_{\rm rot}$, $s_{\rm 0.5}$,
$s_{\rm 1.0}$  or $S$.
Each symbol represents the averaged value over the entire sample of galaxies   
(upper left panel) and the subsamples S1, S2 and S3 defined in section \ref{sec:kinematics}
(upper right panel, lower left panel and lower right panel, respectively).
The corresponding standard deviations are also shown.  Note that the radii are given normalised to
$R_{\rm bar}$ in the main panels while, in the insets, they are given normalised to $R_{\rm max}$.
The horizontal solid line indicates where
$V_{\rm i}$ agrees with $V_{\rm circ}$, while the dashed horizontal lines denote
the range where $V_{\rm i}$ does not depart more than 20\% from $V_{\rm circ}$.
By analysing the main panels of this figure it is clear that $\sigma$ is lower than $V_{\rm circ}$
at almost all analysed radii
with the only exception of the inner parts of the galaxies.  At radii larger
than $0.5 R_{\rm bar}$, $\sigma / V_{\rm circ}$ remains almost constant at
$\approx 0.55$ if we consider the whole sample of galaxies, $\approx 0.50$ in the
case of S1 and $\approx 0.60$ for S2 and S3.
These findings are consistent with the fact that S1 is comprised by  more
disc-like galaxies and hence, the dispersion velocities tend to be smaller.
For the whole galaxy sample, our results indicate that $V_{\rm rot}$ can be considered
a proxy for $V_{\rm circ}$ in the range $0.2 R_{\rm bar}<r<0.5 R_{\rm bar}$.
Similar trends can be appreciated for the subsamples S2 and S3 but, in the case of S1, this range can be extended towards $0.7 R_{\rm bar}$, approximately.
By comparing $V_{\rm circ}$
with $s_{\rm 0.5}$ and $S$, we see that the former underestimates
$V_{\rm circ}$, while the latter overproduces it for the whole sample, and for S1, S2 and S3. 
Nevertheless, in the case of S1, $S$ constitutes a better representation of $V_{\rm circ}$
as a consequence of the lower dispersion velocities in this subsample.
The analysis of the rotation and dispersion
velocity curves of our simulated galaxies suggests that the best proxy for $V_{\rm circ}$
at $z=0$ is  $s_{\rm 1.0}$, which approximates remarkably well the simulated $V_{\rm circ}$ 
at $r>0.5 R_{\rm bar}$ for our whole sample and each one of the subsamples.

When analysing the behaviour of
$V_{\rm i} / V_{\rm circ}$ as a function of $R / R_{\rm max}$ (insets of Fig. \ref{fig:ViVc_vs_r}),
we note that for the whole sample of galaxies the scatter in the vertical axis increases significantly,
mainly at $R / R_{\rm max} \sim 1$.  We can see that the main contribution to this scatter
comes from galaxies in S2 and S3, while in the case of S1 the scatter is quite small.
 For S3, the scatter is even larger than for S2 as these systems exhibit a more disturbed baryonic kinematics 
(misaligned gas discs).  It is interesting to note that, in these simulations, the rotation
curves of spheroid-dominated systems tend to be more affected in the inner parts of the galaxy
($R \sim R_{\rm max}$) than outwards ($R \sim R_{\rm bar}$).
It is also worth noting that for the three subsamples, on average,
$V_{\rm rot} \approx V_{\rm circ}$ at $R \sim R_{\rm max}$.

\begin{figure}
\begin{center}
\hspace{-1cm}\resizebox{9.5cm}{!}{\includegraphics{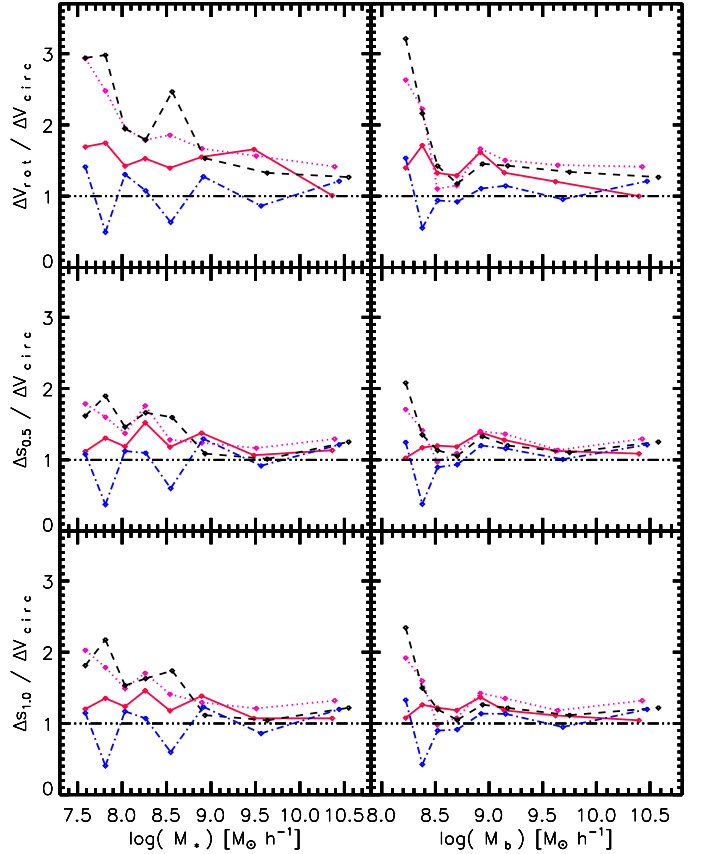}}
\end{center}
\caption[$\Delta V_{\rm i} / \Delta V_{\rm circ}$ at $z=0$]
{
Ratio between the standard deviations of $V_{\rm i}$ and 
$V_{\rm circ}$  ($\Delta V_{\rm i} / \Delta V_{\rm circ}$) at $z=0$ for different bins
of stellar ($M_{*}$) and
baryonic ($M_{\rm b}$) mass.  Each bin is constituted by 35 galaxies of similar stellar or baryonic masses.
Results are shown for $V_{\rm i}$: $V_{\rm rot}$ (upper panels), $s_{\rm 0.5}$ (middle panels)
and $s_{\rm 1.0}$ (lower panels).
The kinematical indicators are estimated by evaluating the gas kinematics at different radii in the simulations:
$R_{\rm max}$ (dot-dashed line), $0.5 R_{\rm bar}$ (solid line), $R_{\rm bar}$ (dotted line)  and $1.5 R_{\rm bar}$ (dashed line).
}
\label{fig:dVidVc_vs_r}
\end{figure}

\begin{figure*}
\begin{center}
\resizebox{6.5cm}{!}{\includegraphics{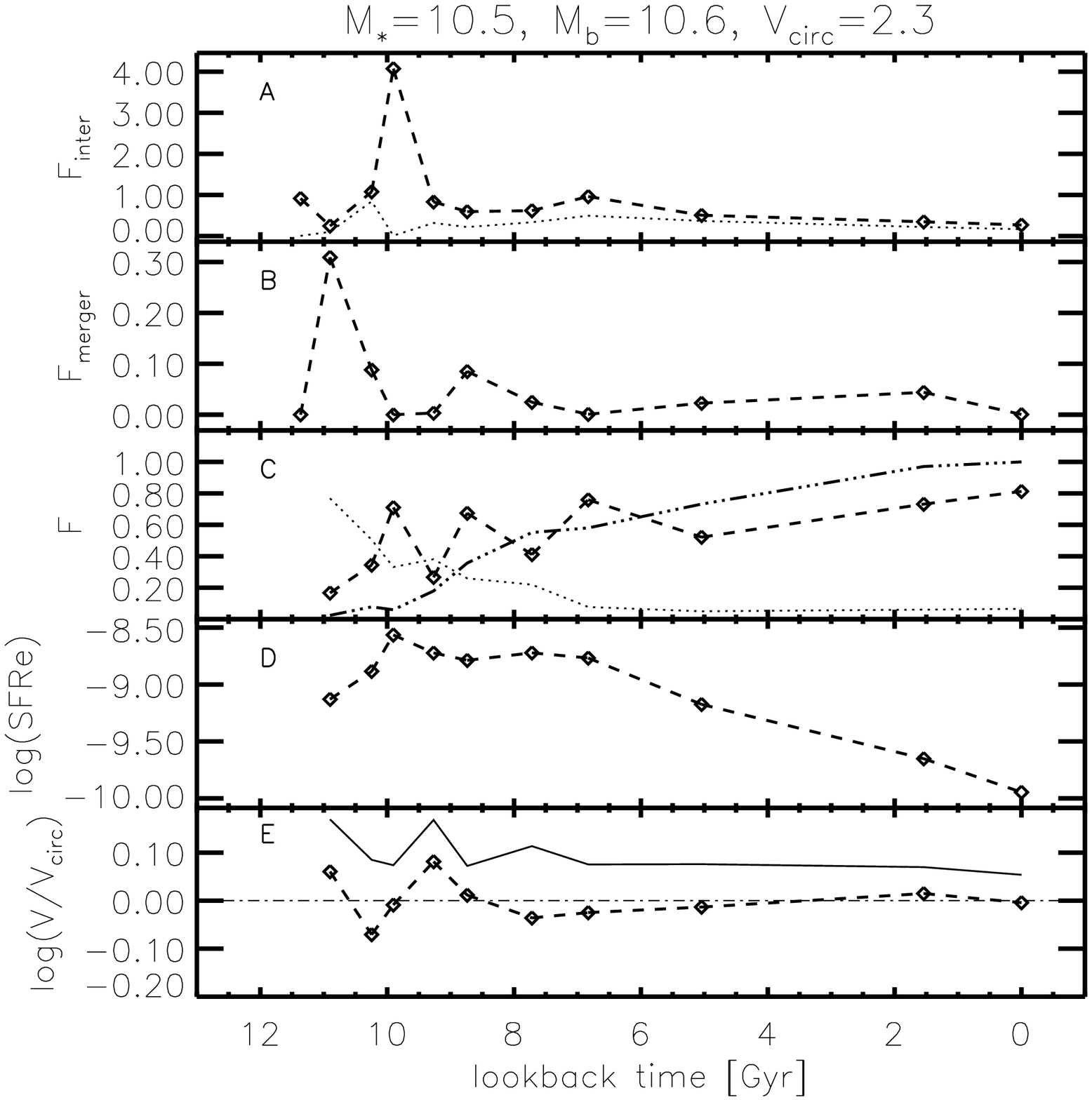}}
\resizebox{6.5cm}{!}{\includegraphics{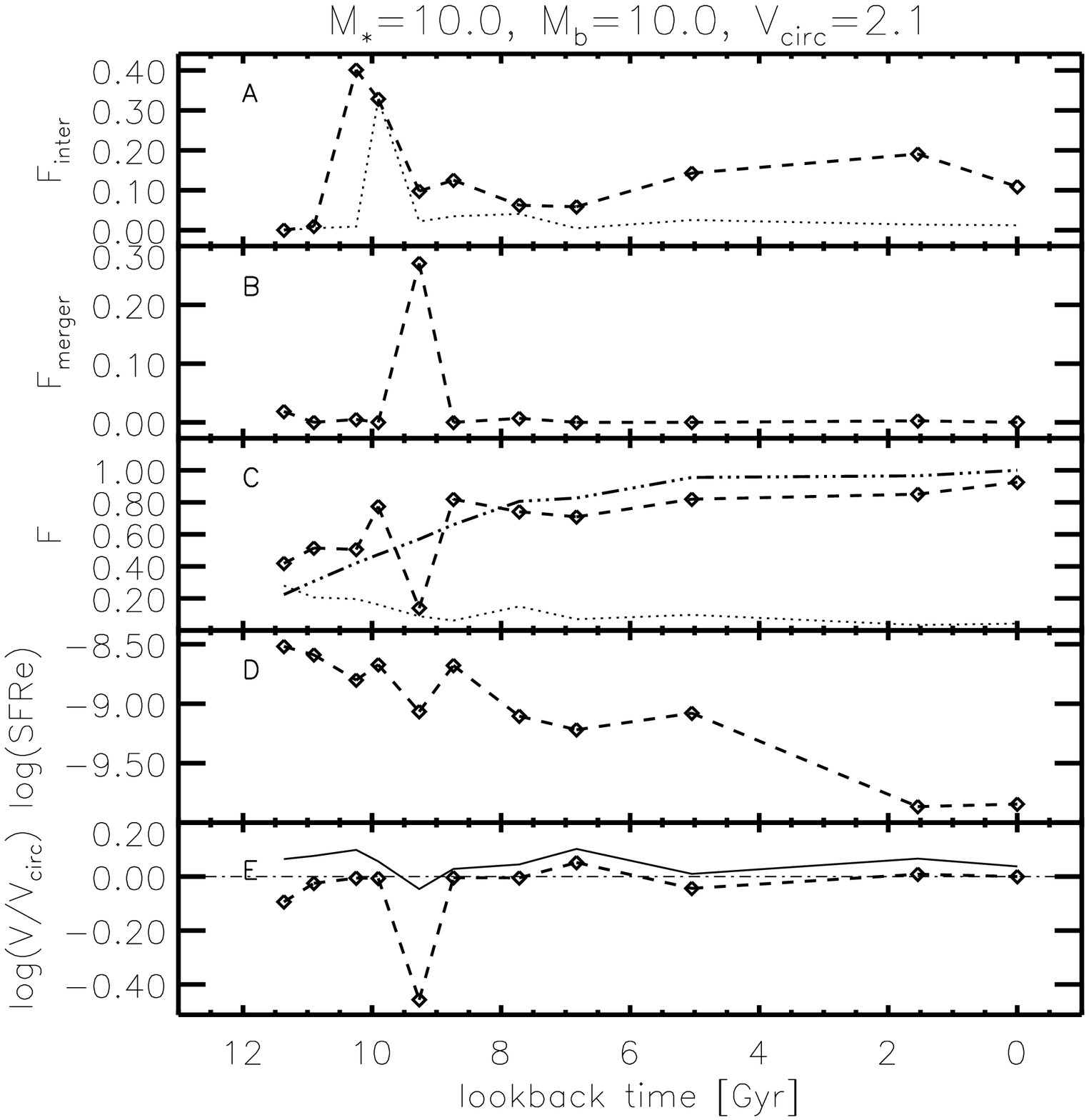}}\\
\resizebox{6.5cm}{!}{\includegraphics{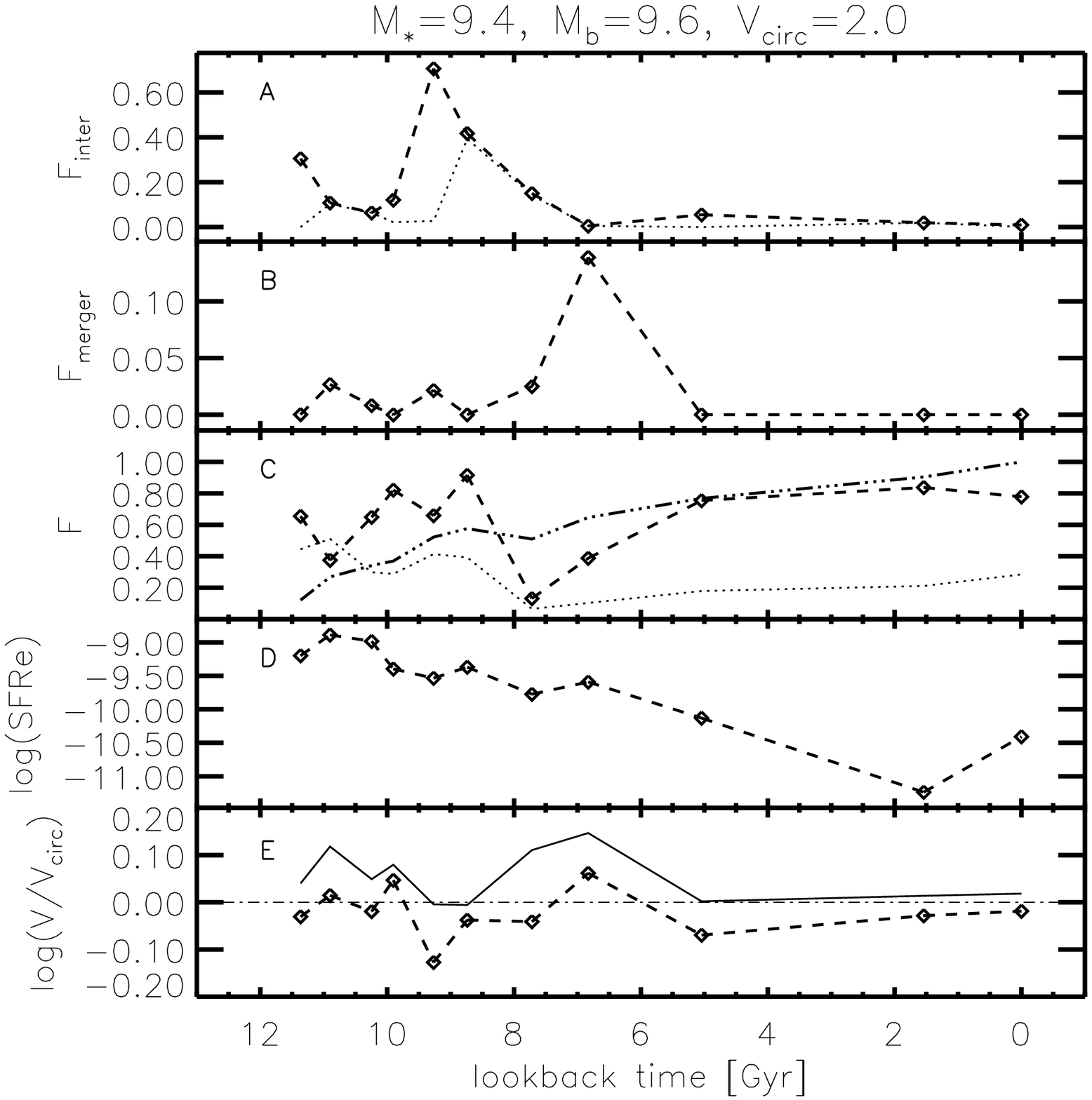}}
\resizebox{6.5cm}{!}{\includegraphics{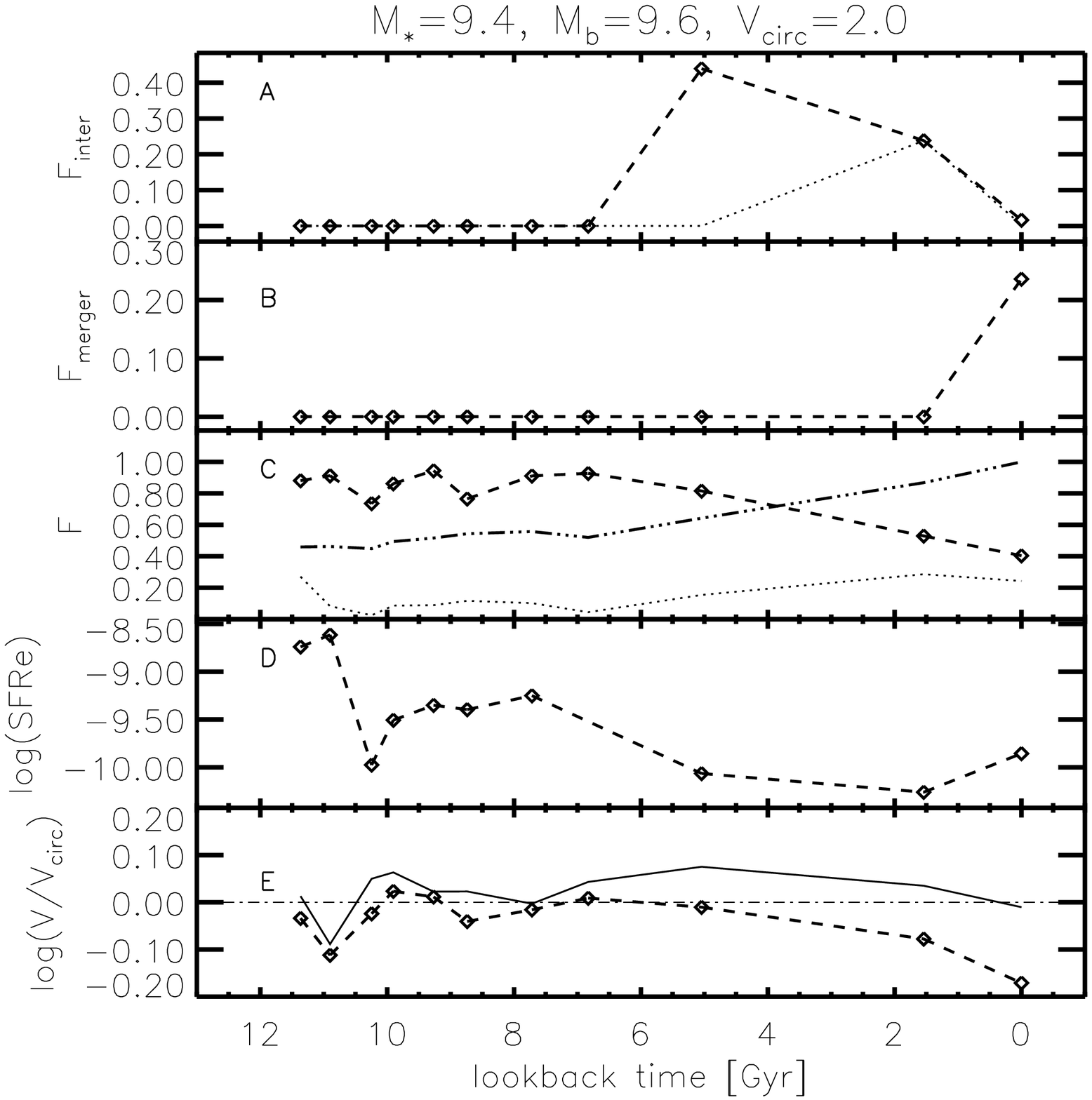}}\\
\resizebox{6.5cm}{!}{\includegraphics{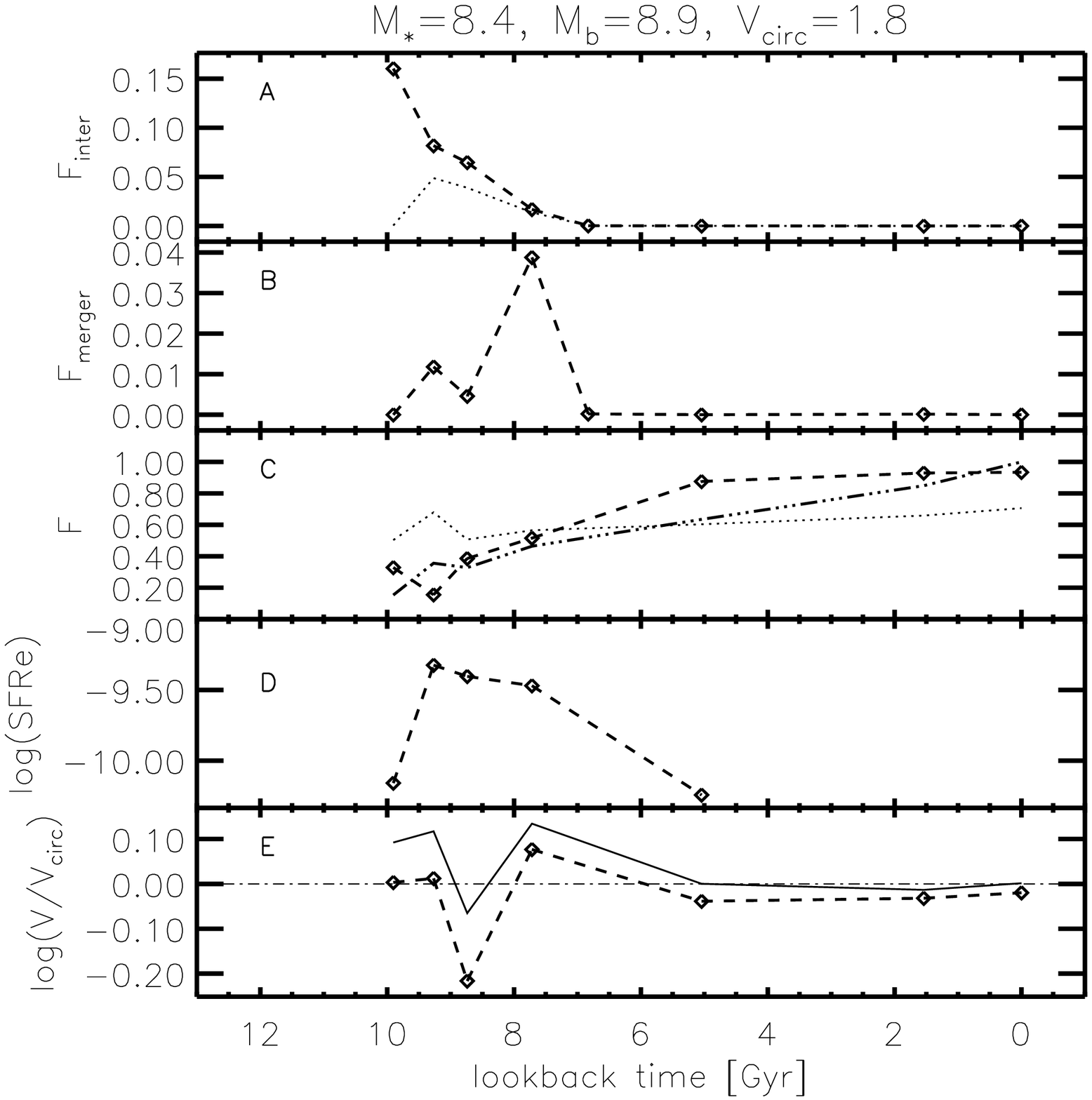}}
\resizebox{6.5cm}{!}{\includegraphics{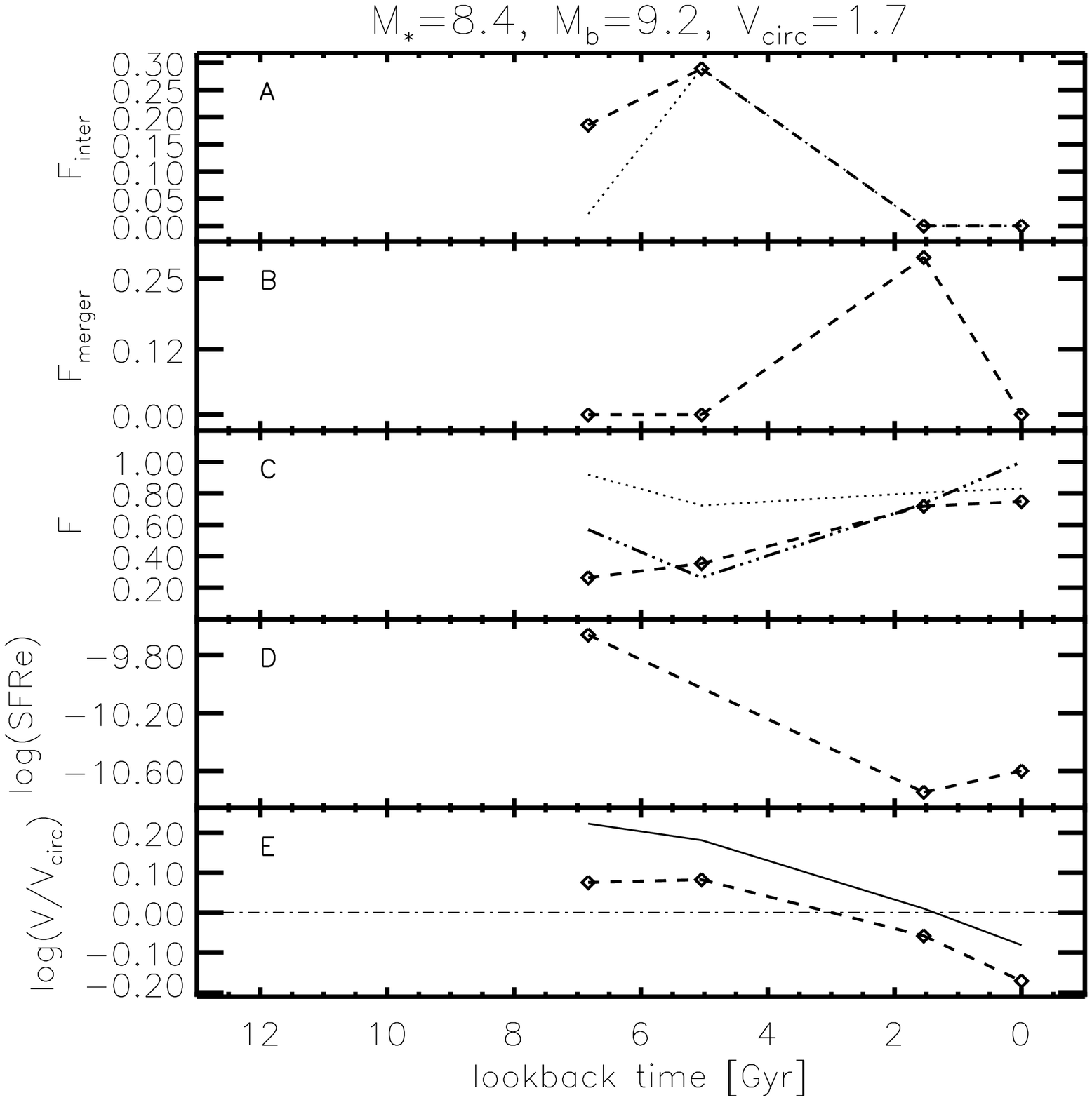}}
\end{center}
\caption[Formation Histories]
{
Formation histories for six typical
simulated galaxies at $z=0$ in the simulations,  with different stellar ($M_*$) and
baryonic ($M_{\rm b}$) masses and circular velocities ($V_{\rm circ}$).
The masses are given in logarithm and in units of $M_{\odot} h^{-1}$
The circular velocities are given in logarithm and in units of ${\rm km \, s}^{-1}$.
In panels A, we show the evolution of $F_{\rm inter}^{R_{\rm vir}}$ (dotted line) and
$F_{\rm inter}^{2 R_{\rm vir}}$ (dashed line). In panels B, we can appreciate 
$F_{\rm merger}$.  Panels C show the time evolution of the gas fraction 
(dotted line), baryonic mass relative to the final one (triple-dot-dashed line) 
and $F_{\rm disc}^{\rm g}$ (dashed line). Panels D show the star formation 
efficiency SFRe in ${\rm yr}^{-1}$ and, in panels E, we can appreciate
the ratio $V_{\rm rot} / V_{\rm circ}$  (dashed lines) and $s_{1.0} / V_{\rm circ}$ (solid lines) evaluated at $R_{\rm bar}$.
See the text for definitions of variables.
}
\label{fig:histform}
\end{figure*}

In Fig. \ref{fig:dVidVc_vs_r}, we plotted the ratio between the standard deviation
of the TFR by using $V_{\rm rot}$, $s_{\rm 0.5}$ and $s_{\rm 1.0}$ and the standard
deviation by using $V_{\rm circ}$.  The ratio is analysed as a function
of stellar and baryonic mass.  We can appreciate that by combining $V_{\rm rot}$
and $\sigma$ in the definition of the kinematical indicator $s_{\rm 0.5}$ or
$s_{\rm 1.0}$, the dispersion in the sTFR and bTFR at a given mass is considerably reduced.
By analysing these trends for S1, S2 and S3 separately, we found that this scatter
is mostly generated by galaxies in S2 and S3  (but mainly in S3) given the lower
mass fractions associated to their surviving discs.
Nevertheless, all kinematical indicators including $V_{\rm rot}$
leads to the tightest relation if evaluated at $R_{\rm max}$.  Hence,
$V_{\rm rot}$ at $R_{\rm max}$ not only seems to be a very good representation
of the potential well of simulated systems, regardless of the morphology 
(Fig. \ref{fig:VrVc_vs_pm}, \ref{fig:ViVc_vs_r}) but also, 
the scatter of the sTFR and bTFR tends to be reduced if the
velocity indicators are evaluated at this radius.
This fact may be suggesting that $R_{\rm max}$ can be considered a characteristic
radius for the gas kinematics of galaxies, at least in the case of the potential wells
reproduced by these simulations.
From Fig. \ref{fig:dVidVc_vs_r}, we also appreciate that
both, $s_{\rm 0.5}$ and $s_{\rm 1.0}$
reduces the scatter of the TFR by similar amounts but, in the literature, most authors use $s_{\rm 0.5}$.
In the case of these simulations, we found  $s_{\rm 1.0}$ to be a better
tracer of the gas kinematics as it is also a good proxy for $V_{\rm circ}$
(Fig. \ref{fig:ViVc_vs_r}).
Finally, by comparing the relative deviations of the sTRF and bTFR (left and right-hand panels of Fig. \ref{fig:dVidVc_vs_r}), 
it is evident that the bTFR is tighter
than the sTFR for the analysed mass range everywhere, but at the lowest masses.  
The less massive systems are also the most affected by numerical resolution so the trends at the low-mass
end should be confirmed with
higher numerical resolution simulations.

\section{The effects of galaxy assembly in the TFR-plane}
\label{sec:evolution}
In order to understand the evolution of the gas kinematics of galaxies 
and its implications for the origin of the scatter in the sTFR and bTFR, we 
analyse the evolution of galaxies
on the TFR-plane as a function of time and in relation to important events such as mergers.
To accomplish this purpose,
we reconstructed the assembly history of the complete sample of simulated galaxies
from their formation time
to $z=0$ and explored which events disturb their rotation curves
leading to outliers on the TFR-plane.
We designed our own algorithm to follow the main progenitor branch for each
simulated galaxy.   At each available time, we define the main progenitor as the most massive
baryonic substructure. All the other smaller systems in the tree are considered satellites which will eventually merge onto the main branch.

In order to evaluate the effects of the galaxy-galaxy interactions, for a given galaxy,
we determined the total mass within $R_{\rm bar}$ 
($M_{\rm dyn}^{\rm main}$) and the mass of  all its neighbours ($M_{\rm dyn}^{\rm neighb}$) located at a distance
smaller than a certain radius $D$, which can be taken either as the virial radius ($R_{\rm vir}$) or as twice its value.
We  quantified the strength of interactions by the ratio 
 $F_{\rm inter}^D = \sum M_{\rm dyn}^{\rm neighb} / M_{\rm dyn}^{\rm main}$. This ratio is measured
 at both distances
in order to have an idea of the satellite distribution with respect to the main progenitor 
(for example if most of them are within  $R_{\rm vir}$  or further away).
To quantify the strength of  mergers, the ratio $F_{\rm merger} = \sum M_{\rm dyn}^{\rm sat} / M_{\rm dyn}^{\rm main}$
is defined, where  $M_{\rm dyn}^{\rm sat}$ is the mass of the satellites which were accreted between two available
times. 

For the main progenitor, we  estimated the 
gas fraction, its baryonic mass normalised to the value at $z=0$ 
and $F_{\rm disc}^{\rm g}$ as a function of lookback time in order to quantify the building up of 
the galaxy as well as the
role of outflows, as we will discuss in next section. We also calculate the star formation rate efficiency (SFRe) in the progenitor galaxy as the
ratio between the star formation rate and the gas mass.
We will use the ratios $V_{\rm rot} / V_{\rm circ}$ and $s_{1.0} / V_{\rm circ}$ to quantify the disturbance of the rotation curve along the evolutionary paths.
All these quantities will be estimated at $R_{\rm bar}$.

\subsection{Formation histories of typical galaxies}
\label{sec:histform}

We analysed the evolution of the quantities defined above for the main
progenitor branch of the local simulated galaxies as a function of lookback time.
In order to illustrate the relation between interactions and mergers and disturbances of the rotation curve, we selected
six typical galaxies out of a sample of 309 galaxies at $z=0$, as shown in Fig. \ref{fig:histform}.

The galaxy in the upper-left panel of Fig. \ref{fig:histform} is one of the more  massive galaxies at $z=0$  
exhibiting a lower scatter in the local simulated TFR.  Note that, as expected,
$V_{\rm rot} / V_{\rm circ}$ is approximately 1 at $z \sim 0$ while,
$s_{1.0} / V_{\rm circ}$ tends to be a little larger by $\sim 0.1$ dex.  
By analysing its formation history, we see that the main progenitor
 has not  experienced important 
interactions or merger events during the last 5 Gyr.  
Its SFRe has also decreased by more than an order of magnitude in the same time interval
as a consequence of the decrease of the gas fraction.  
Therefore, the galaxy at $z=0$ has reached a passive stage of evolution and stability which 
is consistent with its well-defined rotation curve and low scatter in the TFR.
Note also that, although the gas fraction is small ($\sim 0.07$) at $z=0$, the gas mass within
$R_{\rm bar}$ is $\approx 3 \times 10^9 M_{\odot} h^{-1}$ so that it is possible to 
have a well resolved disc component.
By exploring the main progenitor branch at early times, we can see that 
for a lookback time greater than 8 Gyr the main progenitor exhibited a disturbed kinematics.
At this epoch, the ratio $V_{\rm rot}/V_{\rm circ}$ shows variations by $\pm 0.1$ dex 
over time intervals of around 1 Gyr.  
From the inspection of $F_{\rm inter}$ and $F_{\rm merger}$ at that time,
we can infer that this behaviour can be  associated with
interaction and merger events which lead to a high SFRe and exhausted the gas
reservoir of the system.
It is interesting that merger and interactions can generate either an increase or
decrease of $V_{\rm rot}/V_{\rm circ}$ \citep{pedrosa2008}. 
With respect to $s_{1.0}/V_{\rm circ}$,  we obtained similar trends
but with smaller and more smooth variations.

The galaxy in the upper right panel is an example of an intermediate-mass
galaxy with low scatter in the local TFR.  This galaxy also exhibits a passive
evolution within the last few  Gyrs. Similarly to the previous example, we can see that at early times ($ \sim 9 $ Gyrs ago), its
 main progenitor experienced important interactions followed by  merger events 
($F_{\rm inter}^{2 R_{\rm vir}} \sim 0.4$, $F_{\rm inter}^{R_{\rm vir}} \sim 0.3$, $F_{\rm merger} \sim 0.25$) which could
be linked to a decrease in $V_{\rm rot}/V_{\rm circ}$ ratio
by around $\sim 0.5$ dex over the corresponding period.  Note also that $F^{\rm g}_{\rm disc}$ exhibits a minimum at
that epoch.  After these events, its disc component was
reconstructed and the gas rotation velocity approached the circular velocity again.
It is worth noting that $s_{1.0}/V_{\rm circ} \sim 1$ during the whole formation history of this
galaxy, even during the epoch of mergers an interactions.  As previously reported in the literature,
these findings suggest that the combination of $V_{\rm rot}$ and $\sigma$ in the definition of the
kinematical indicator generates a velocity scale more robust against disturbances of the gas kinematics.

In the middle panels of Fig. \ref{fig:histform}, we compare the formation histories
of two smaller galaxies with similar stellar masses of $ M_{*} \approx 10^{9.4}   M_{\odot} h^{-1}$
and circular velocities of $V_{\rm circ}  \approx 100  \ {\rm km \, s}^{-1}$.  The galaxy in the left panel
exhibits a  smaller scatter in the local TFR than  the galaxy in the right one. 
  By comparing the recent formation histories of
these galaxies,  we see that  the large scatter in the second one can  be associated to  
 strong interactions ($F_{\rm inter}^{2 R_{\rm vir}} \sim 0.4$ and 
$F_{\rm inter}^{R_{\rm vir}} \sim 0.2$) taking place during the last 6 Gyrs and which ended up 
in a  recent important merger event ($F_{\rm merger} \sim 0.25$).
Before this epoch, the $V_{\rm rot}/V_{\rm circ}$ for
the main progenitor  remains close to $\sim 1$.  
In the case of $s_{\rm 1.0}/V_{\rm circ}$, the behaviour is similar but with more
smooth variations.
Note that simultaneously with the  interactions and merger events, 
 the gas fraction increases and the percentage of gas in the disc component decreases 
indicating the presence of gas inflows which
contribute more importantly to the formation of a spheroidal component.
In the case of the galaxy in the left panel, the lack of mergers and interactions
during the last 5 Gyr is in agreement with the small variations of $V_{\rm rot}/V_{\rm circ}$
and the high values of $F_{\rm disc}^{\rm g}$.
Note, however, that its main progenitor  experiences an increase of $V_{\rm rot}/V_{\rm circ}$
7 Gyrs ago, during a merger event ($F_{\rm merger} \sim 0.15$), while  9 Gyr ago,
a strong interaction ($F_{\rm inter}^{2 R_{\rm vir}} \sim 0.6$) can be linked to a decrease
by more than 0.1 dex in $V_{\rm rot}/V_{\rm circ}$. 
Once again, 
the trends for $s_{\rm 1.0}/V_{\rm circ}$ are similar to those appreciated for $V_{\rm rot}/V_{\rm circ}$ but with a more smooth evolution.
Note that, as seen before,  $F_{\rm disc}^{\rm g}$
reaches a minimum during the period of strong interactions.

Finally, in the lower panels,  we compare the evolution of galaxies
with low stellar masses of $M_{*}  \approx 10^{8.4} M_{\odot} h^{-1}$
and circular velocities of $V_{\rm circ}  \approx 50 \ {\rm km \, s}^{-1}$.  The galaxy in the left panel
shows a smaller scatter in the local TFR than the  one in the right panel.
Even in these small galactic systems, the scatter in $V_{\rm rot}/V_{\rm circ}$ 
seems to be related to the presence of merger events.   
The galaxy in the left panel has not suffered significant mergers or
interactions during the last 5 Gyrs in agreement with its small changes of $V_{\rm rot}/V_{\rm circ}$.
Note, however, that the progenitor in the main branch has been affected by mergers and interactions 8 Gyr
ago, which led to a peak in the  SFRe and caused changes by $\sim 0.25$ dex in  $V_{\rm rot}/V_{\rm circ}$
and $< 0.15$ dex in the case of $s_{\rm 1.0}/V_{\rm circ}$.
For the galaxy in the right panel,  the distortion of the rotation curve at $z \sim 0$ 
can be  again related with recent important mergers and interactions: $F_{\rm inter}^{R_{\rm vir}} \sim 0.3$ and  $F_{\rm merger} \sim 0.3$.

Therefore, our results suggest that the hierarchical aggregation of
the structure strongly influences the evolution of the galactic gas component within a $\Lambda$-CDM Universe,
producing TFR outliers in concordance with previous works 
\citep[e.g.][]{simard1998, barton2001, kannappan2004, 
bohm2004, flores2006, weiner2006, kassin2007}.
Mergers and interactions seem to be important at regulating the gas kinematics
and star formation process for the whole range of stellar masses covered by
these simulations, affecting the evolutionary tracks of systems on the TFR-plane.
All these phenomena seem to be responsible for the scatter we found in the TFR.
As previously reported in the literature \citep[e.g.][]{kassin2007}, we obtained that the combination of
$V_{\rm rot}$ and $\sigma$ leads to a velocity scale more stable during merger
and interaction events reducing the scatter in the sTFR and bTFR.  
We obtained, for example, that $|\log (V_{\rm rot} / V_{\rm circ})|<0.5$ for
the whole sample of simulated systems while $|\log (s_{\rm 1.0} / V_{\rm circ})|<0.3$.
Note, however, that some observational works reported
scatters in the sTFR greater than $\sim 1$ dex \citep[e.g.][]{kassin2007}.  
In this context, it is worth
mentioning that the dispersion in the observed TFR relation might depend very sensitively
on the observational approach and modelling techniques \citep[e.g.][]{miller2011} and it is yet
not clear which is the intrinsic scatter of the relation.
By analysing numerical simulations, \citet{covington2010} found scatters larger than $1$ dex.
Nevertheless, those simulations
were pre-prepared mergers of two isolated galaxies.  
In this work, by studying cosmological simulations, we are able to follow the
evolutionary history of galaxies in a self-consistent way.

\subsection{Fingerprints of formation histories on the TFR-plane}

In this section, we  investigate how the assembly histories 
of galaxies in a cosmological scenario
induced changes  in the TFR-plane searching for fingerprints of particular events
which, then, can be used to understand observations.

\begin{figure*}
\begin{center}
\resizebox{6.5cm}{!}{\includegraphics{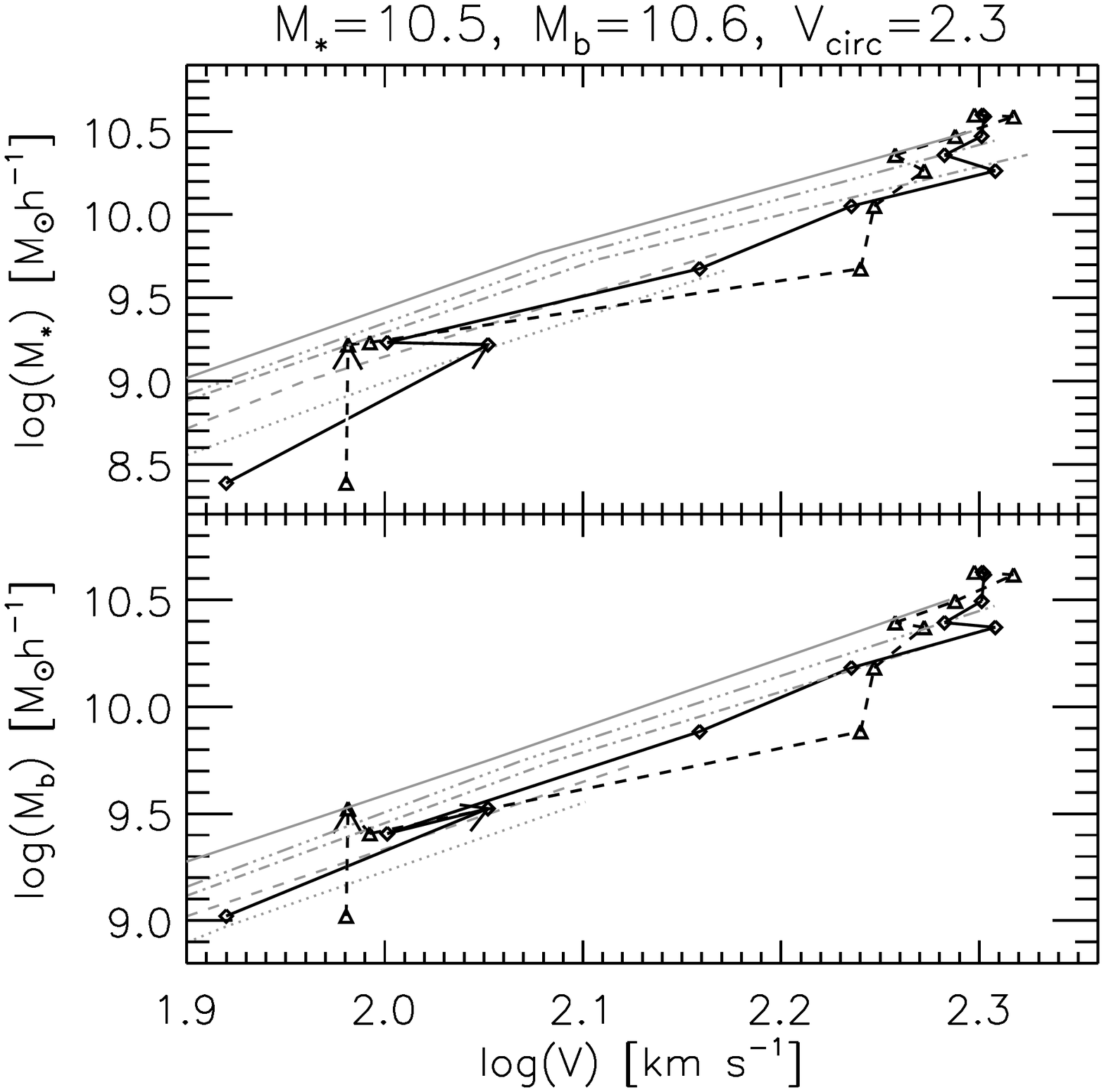}}
\resizebox{6.5cm}{!}{\includegraphics{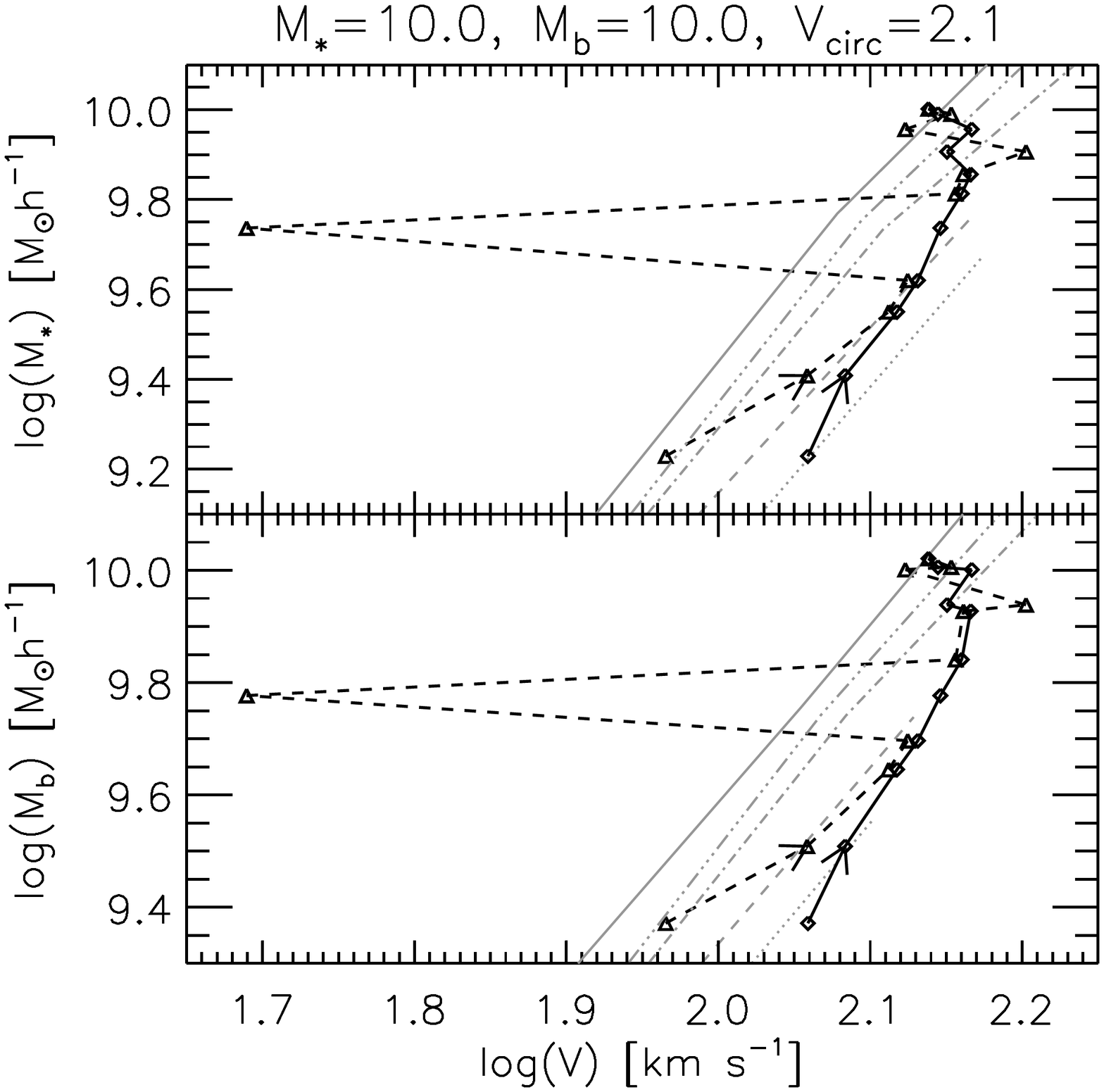}}\\
\resizebox{6.5cm}{!}{\includegraphics{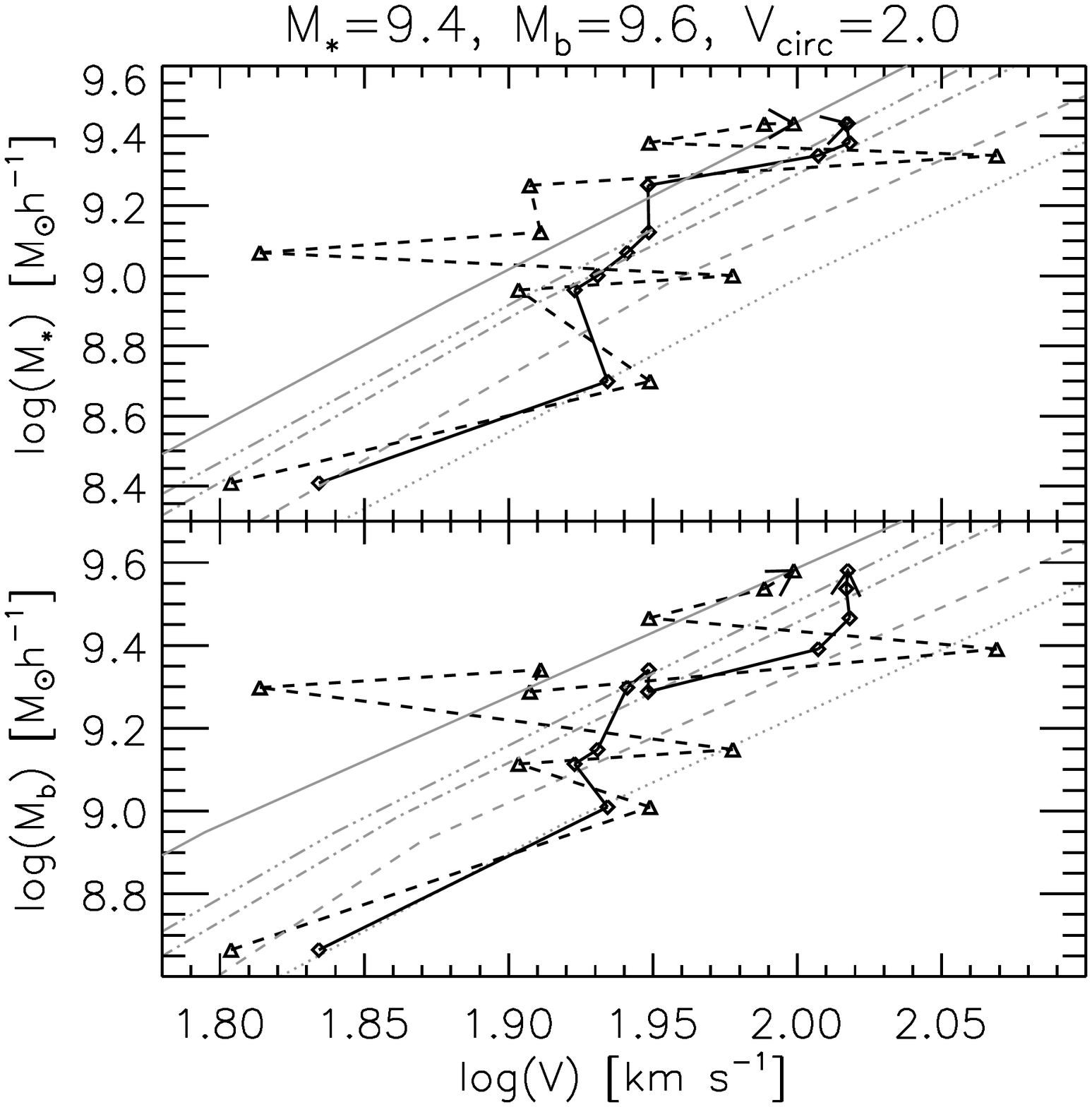}}
\resizebox{6.5cm}{!}{\includegraphics{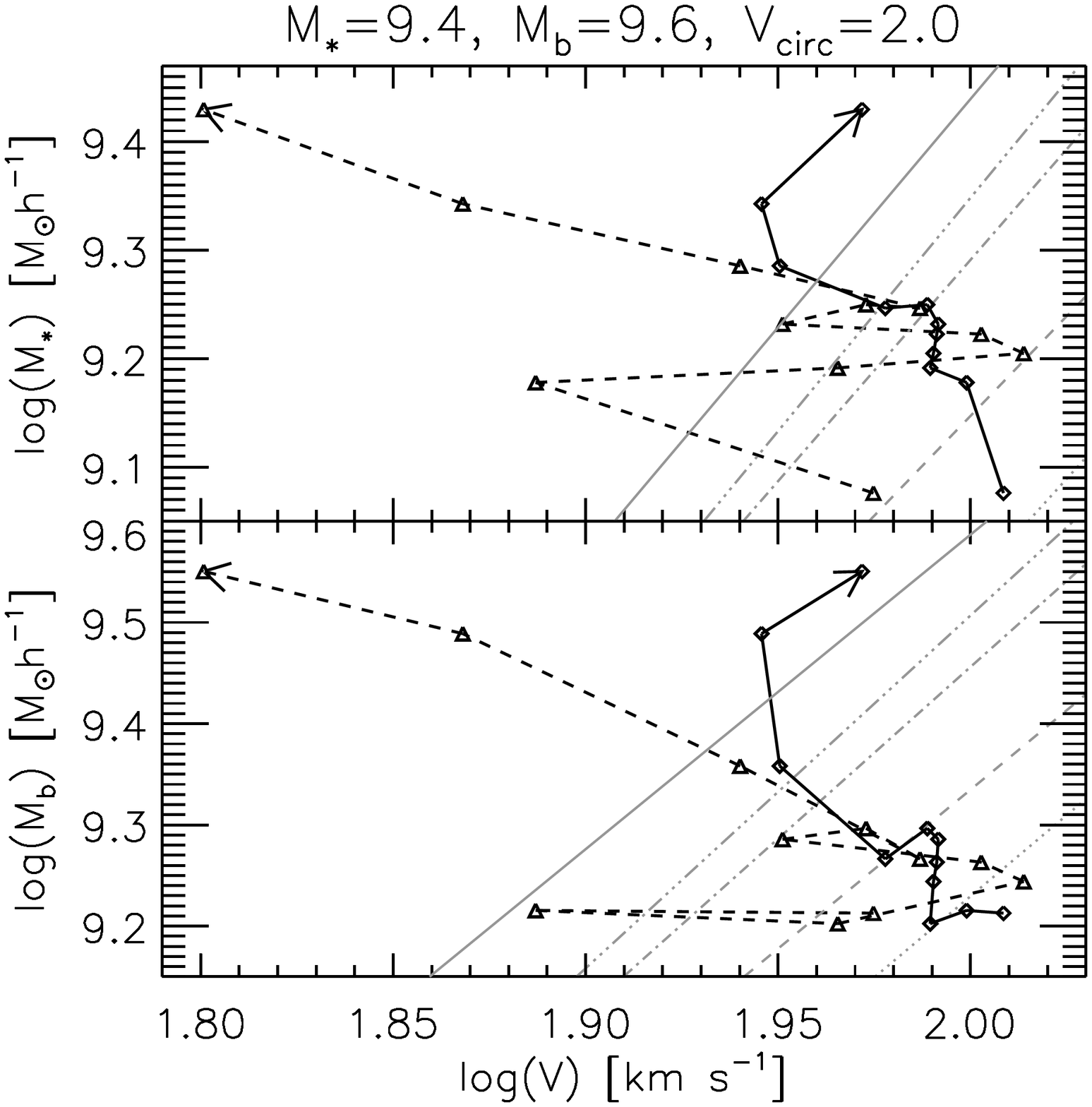}}\\
\resizebox{6.5cm}{!}{\includegraphics{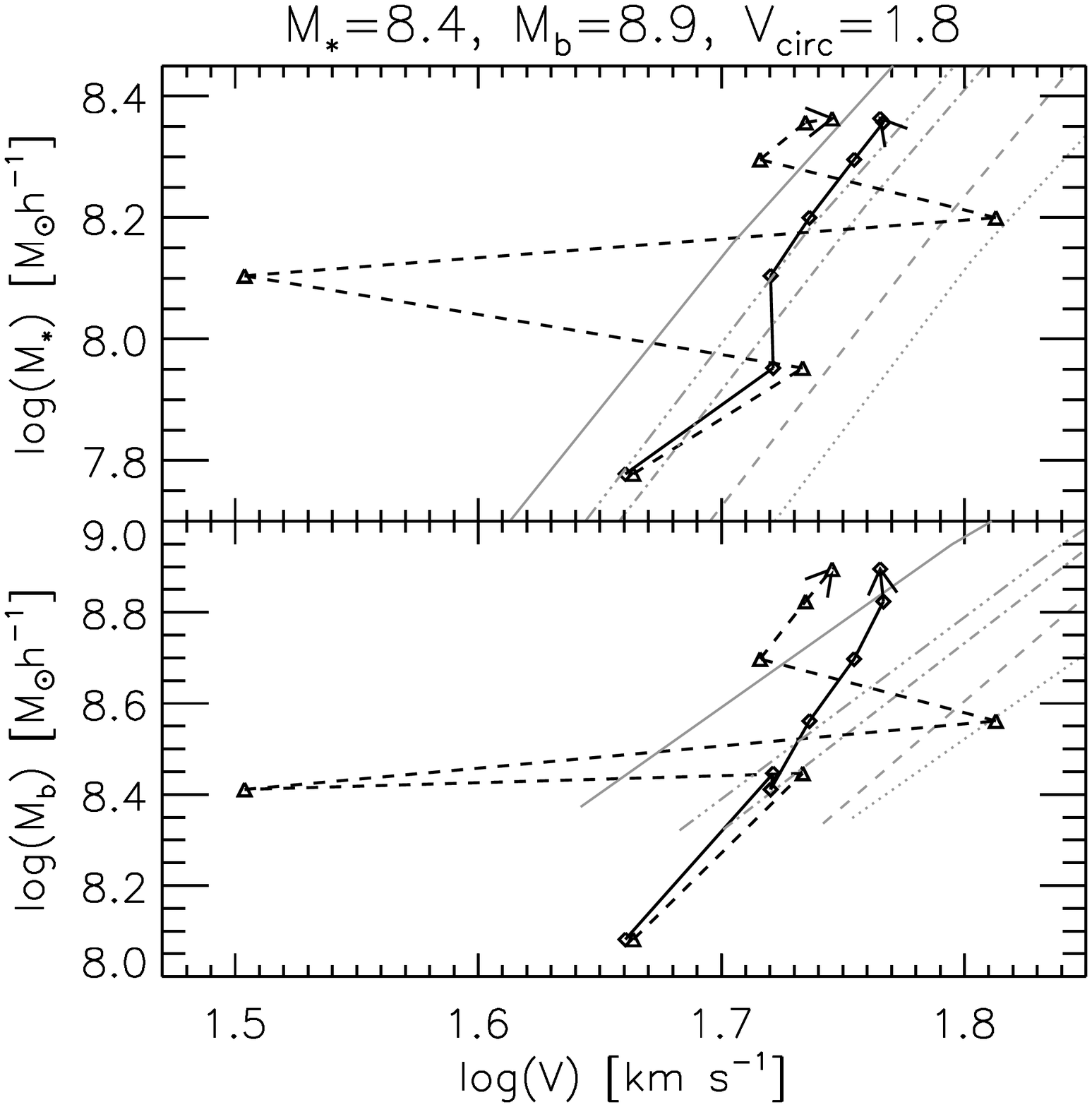}}
\resizebox{6.5cm}{!}{\includegraphics{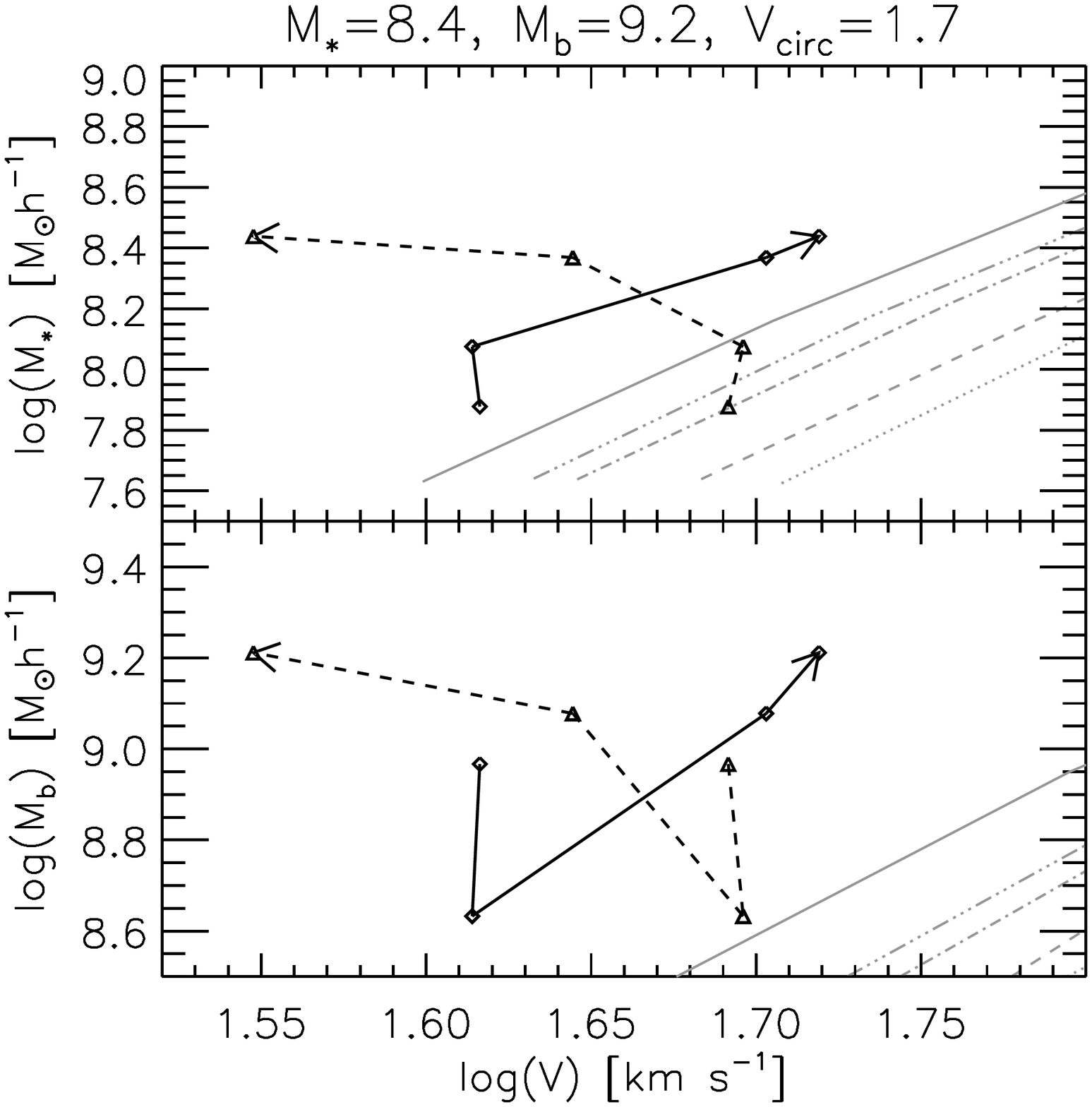}}
\end{center}
\caption[Evolution on the TF-plane]
{
Evolutionary tracks of the six galaxies
shown in Fig. \ref{fig:histform} on the sTFR and bTFR planes. The background grey lines depict the mean sTFR
and bTFR obtained by using $V_{\rm circ}$ at $R_{\rm bar}$ as a kinematical indicator
at $z=3, 2, 1, 0.4, 0$ (dotted, dashed, dot-dashed, triple-dot-dashed, solid line, respectively).
The black lines indicate the track of the given galaxy on the TFR plane 
by using $V_{\rm circ}$ (solid line) and $V_{\rm rot}$ (dashed line) at $R_{\rm bar}$ as the kinematical
indicators.  In each panel, there is an arrow indicating in which direction the galaxy evolves along
each path.
}
\label{fig:TFRglo}
\end{figure*}

In Fig. \ref{fig:TFRglo}, we can appreciate the evolution  on the sTFR and bTFR planes 
of the six galaxies described in the previous section. 
If the  TFR is constructed by using  $V_{\rm circ}$, it provides us with 
 information about how the stellar and baryonic masses assembled within the potential well
of the dark haloes hosting the galaxies, while if 
$V_{\rm rot}$ is used, it gives information related to the  gas kinematics.
To assess when galaxies are outliers of the mean TFR at different redshift,  we also include the mean  sTFR and bTFR of the whole population 
as a function of redshift taken from \citet{derossi2010}.
For the sake of clarity, we did not include 
in Fig. \ref{fig:TFRglo}
the tracks given by  $s_{\rm 1.0}$ 
but they can be easily derived from previous section: 
$s_{\rm 1.0}$ would in general lead to
similar trends to those obtained from $V_{\rm rot}$
but with a less noisier evolution and,
given that $s_{\rm 1.0}$ is larger than $V_{\rm rot}$ by definition, the tracks would be also displaced
towards larger velocities.

By comparing Fig. \ref{fig:histform} and Fig. \ref{fig:TFRglo},
we can understand the origin of the changes as the systems evolve.
In particular, we see that mergers lead to an
increase of $V_{\rm circ}$ and mass in different amounts. 
With respect to interactions, in some cases,
we detect a decrease in $V_{\rm circ}$ due to interactions while in others
the interaction generates an increase of the stellar and baryonic
mass with an almost constant $V_{\rm circ}$.  
When following the TFR tracks
using $V_{\rm rot}$ we found a very noisy evolution which is affected considerably
by interactions and mergers that produce outliers of the TFR.
As expected, we see in Fig. \ref{fig:TFRglo} that the tracks given by $V_{\rm circ}$ are smoother
than those given by $V_{\rm rot}$.  
Most mergers and interactions generate a scatter in the TFR greater than
the level of evolution that we estimated since $z = 3$ for the relation based on $V_{\rm circ}$
in these simulations.
As we noted before, these events can shift the $V_{\rm rot}$ towards lower
or greater values than $V_{\rm circ}$.
These findings suggest that, from the observational point of view,
selection effects and noise are likely to mask the actual evolution of the TFR.

In the case of the galaxy corresponding to the upper left panel, the most
important evolution in the TFR-plane took place more than 8 Gyr ago, during an
epoch of important interactions and mergers 
($F_{\rm inter}^{2 R_{\rm vir}} \sim 4$, $F_{\rm merger} \sim 0.3$)
which generate an increase by more than 1.5 dex in stellar and baryonic mass and
by $\sim 0.4$ dex in circular velocity. 

In the case of the galaxy in the upper right panel, the track given by $V_{\rm rot}$ approximates
the one given by $V_{\rm circ}$ with the only exception of the point associated to a lookback time
of $\sim 9$ Gyr when the system experienced an important merger event ($F_{\rm merger} \sim 0.25$).
As a consequence of the merger, $V_{\rm rot}$ decreases by $\sim 0.4$ dex departing from $V_{\rm circ}$  and generating
a TFR outlier.  After $\sim 1$ Gyr, the disc component was reconstructed,  with $V_{\rm rot}$ approaching $V_{\rm circ}$.

With respect to the galaxies in the middle panels, we see that the evolutionary tracks
given by $V_{\rm rot}$ are noisier than those of  more massive galaxies.  
This suggests that interactions and mergers in these simulations are
more efficient at producing instabilities in the disc component of systems
with shallower potential wells.
In these galaxies, the variations in the rotation velocity 
between two time steps of the simulation are, in some cases, larger 
 that the total level of evolution 
of the TFR based in $V_{\rm circ}$ for $3<z<0$.  In particular, the galaxy in the middle right panel
reaches $z \sim 0$ as an outlier of the TFR as a consequence of recent important interactions followed by a merger event 
($F_{\rm inter}^{2 R_{\rm vir}} \sim 0.4$, $F_{\rm inter}^{R_{\rm vir}} \sim 0.2$, $F_{\rm merger} \sim 0.25$).  
In this figure, it is evident that during mergers and interactions the tracks
given by $V_{\rm circ}$ and $V_{\rm rot}$ may diverge.

In the lower panels of Fig. \ref{fig:TFRglo}, we can appreciate that low-mass galaxies also
describe a similar behaviour.
The galaxy in the left panel becomes an outlier of the TFR given by $V_{\rm rot}$ at a lookback time of $\sim 9$ Gyr due
to interactions and mergers.  After $\sim 1$ Gyr, this galaxy recovered its rotational equilibrium with $V_{\rm rot}$
approaching $V_{\rm circ}$ again.  In the case of the
young galaxy in the right panel, the presence of mergers and interactions displaced
 this galaxy away from the mean TFR based in $V_{\rm circ}$.  We can also appreciate that due to a recent merger event, $V_{\rm circ}$ and $V_{\rm rot}$
evolve in opposite ways.

Although we have pointed out only the correlation between mergers and interactions, from
the previous section, we know that outflows and inflows can also be associated to
sharp changes on the TFR-plane. However, in most cases they also occurred close to
mergers or interactions as a second order effect since these violent events
have been proven to be efficient at triggering gas inflows and starbursts
\citep[which can drive outflows, e.g.][]{barnes1996, mihos1996, tissera2000}.

\subsection{Statistical analysis}
Our results for these six galaxies suggest that the hierarchical building up of
the structure strongly influences the evolutionary paths of galaxies
in the TFR-plane modulating the evolution of its scatter with cosmic time.
In order to generalise these findings,
we performed a statistical study by extending the previous analysis
to the whole sample of simulated
galaxies.
Within the hierarchical aggregation paradigm, galaxies are affected by 
different kind of physical processes such as mergers, interactions, star formation,
outflows, gas accretions, etc.
All these processes act together to drive galaxy evolution
and hence it is not easy to disentangle the effect of each individual
event on the variation of the properties of galaxies.  
It is also possible that the combined actions of two or more 
different astrophysical processes compensate each other to generate a insignificant change 
of the mean properties of the system.
Therefore,  we decided to focus our attention on galaxies which have experienced
extreme physical events
in the very recent past (in the previous 2 Gyr of galaxies
selected at a given $z$). For these cases, we expect   
the evolution of the mean properties of galaxies to be mainly 
affected by a particular extreme physical event.

\begin{figure*}
\begin{center}
\resizebox{8.5cm}{!}{\includegraphics{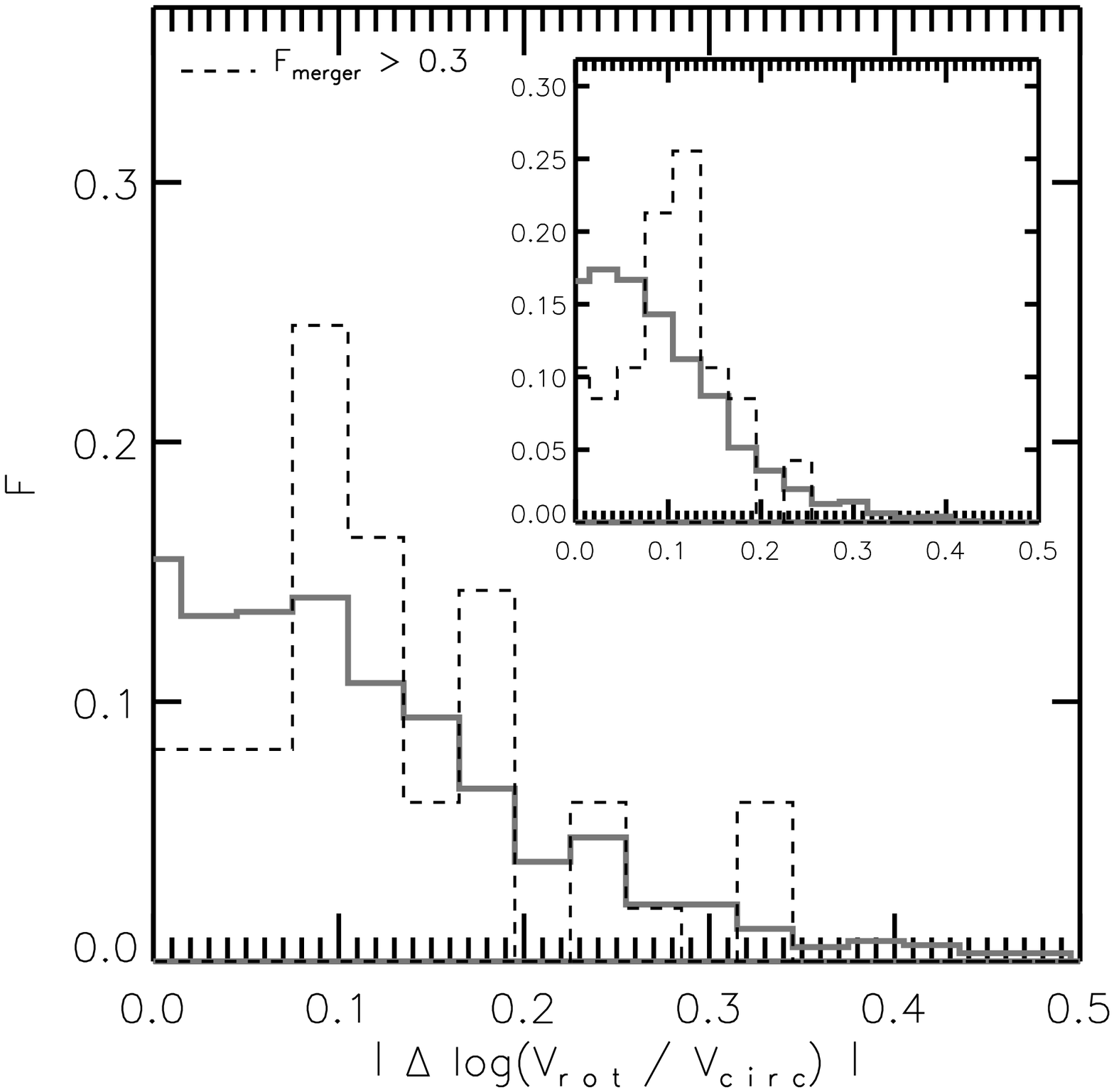}}
\hspace{0.5cm}\resizebox{8.5cm}{!}{\includegraphics{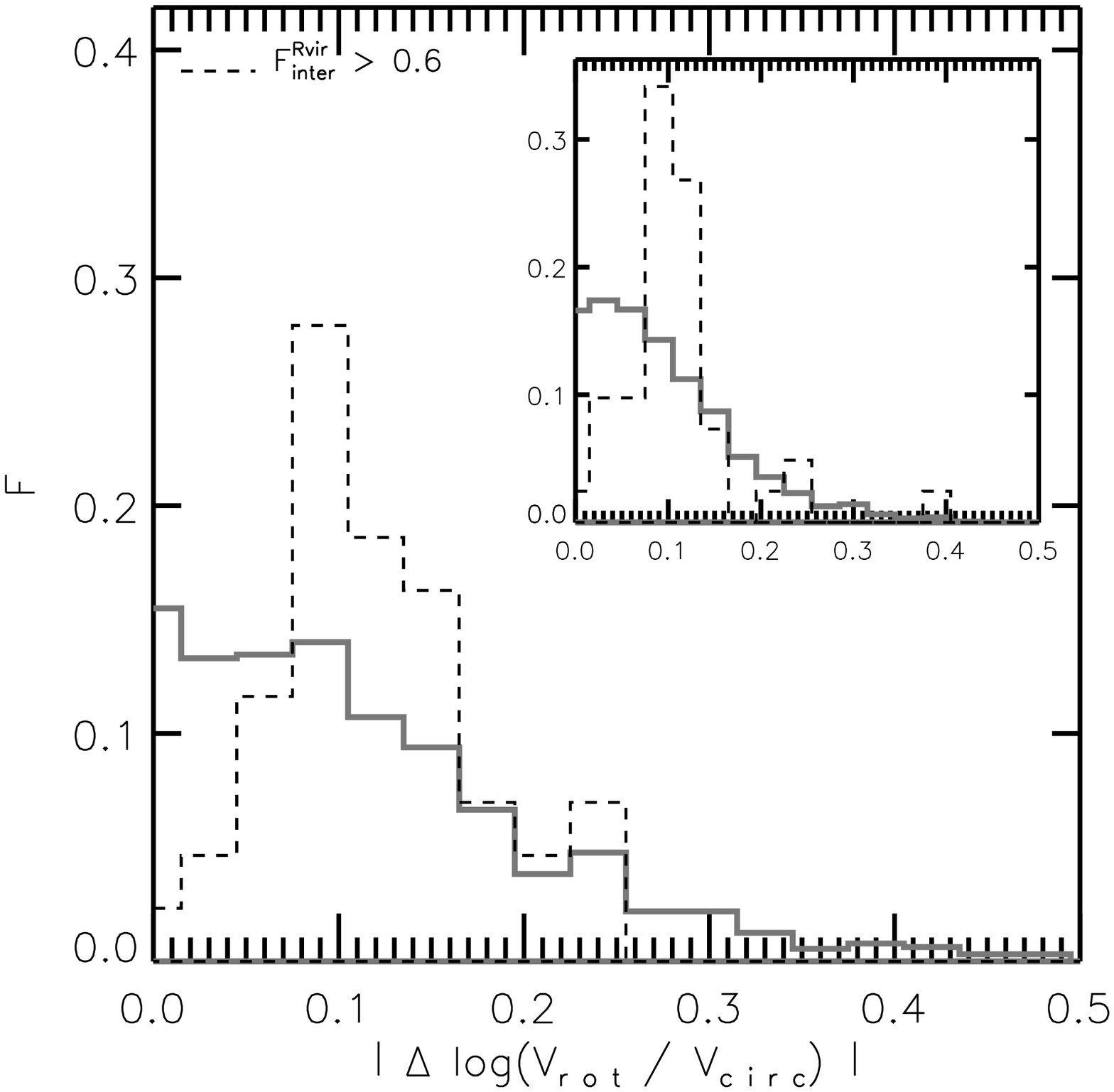}}\\
\resizebox{8.5cm}{!}{\includegraphics{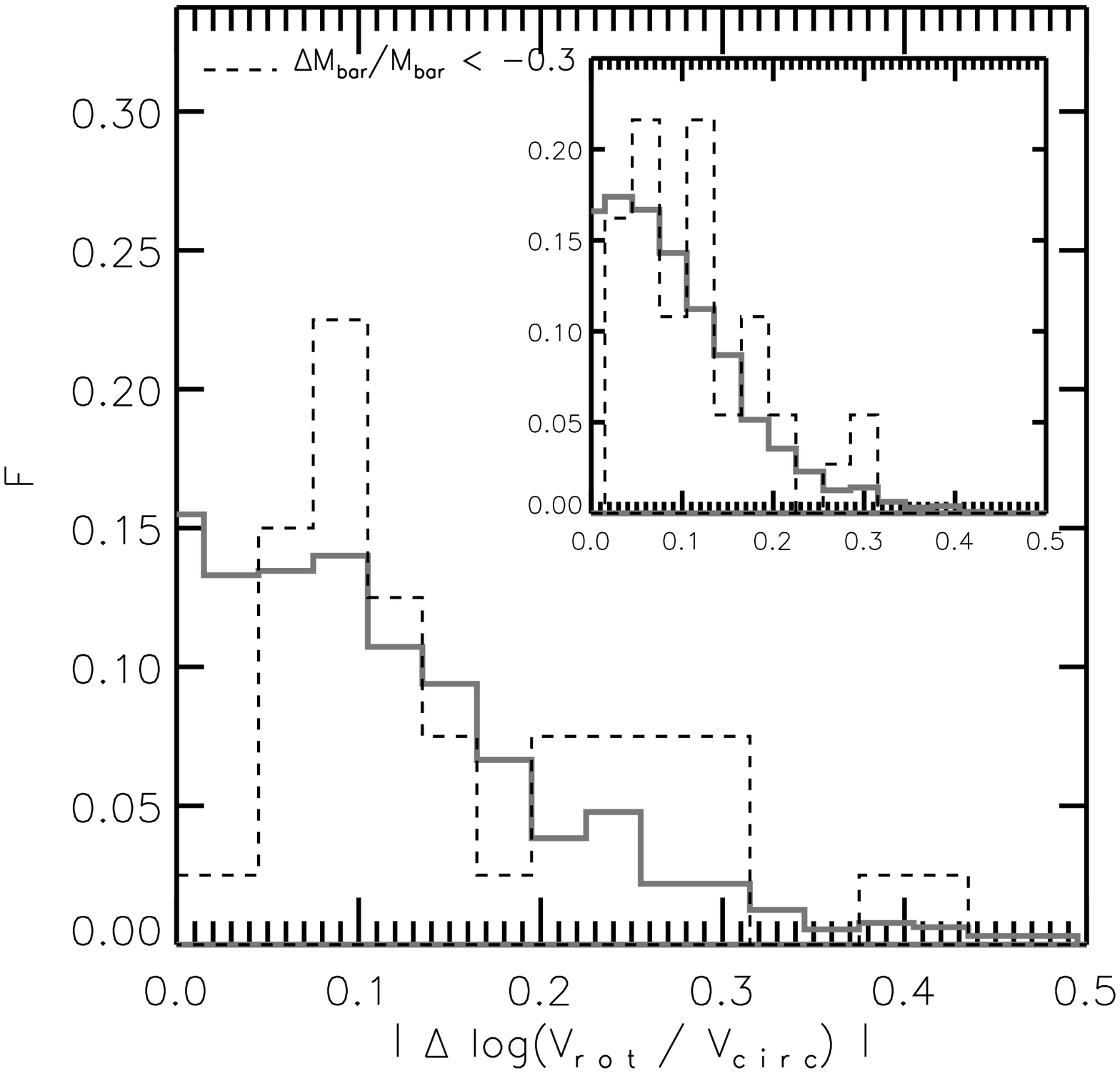}}
\hspace{0.5cm}\resizebox{8.5cm}{!}{\includegraphics{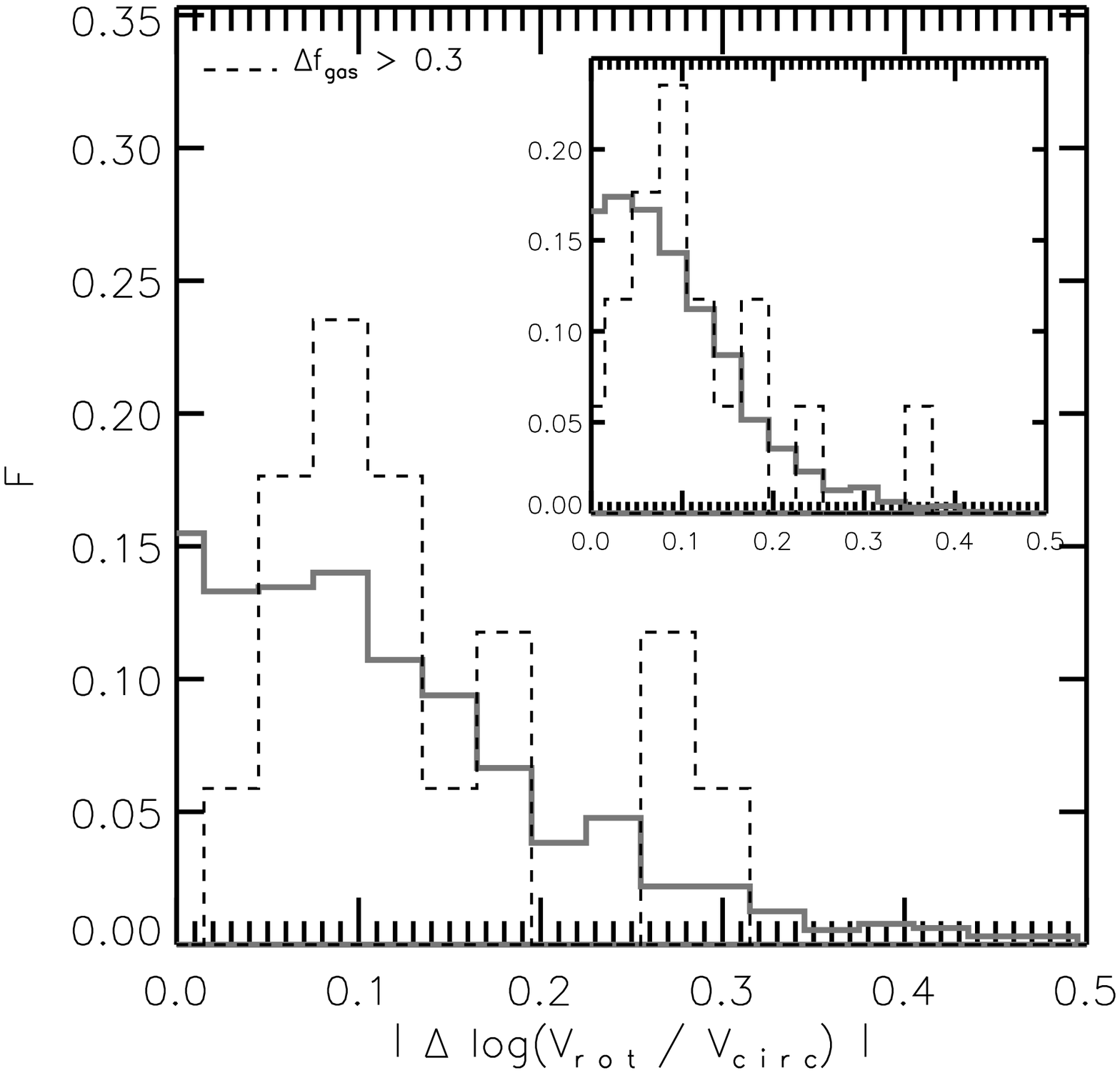}}
\end{center}
\caption[Histograms for different events during galaxy evolution]
{
Distribution of the absolute variation of  $\log (V_{\rm rot} / V_{\rm circ})$ (main panels)
and $\log (s_{\rm 1.0} / V_{\rm circ})$ (insets)
during the last $2$ Gyr of evolution for galaxies which
have been subject to extreme processes (dashed lines):
$F_{\rm merger} > 0.3$ (upper left panel), $F_{\rm inter}^{R_{\rm vir}}  > 0.6$ (upper right panel),
$\Delta M_{\rm bar} / M_{\rm bar} < -0.3$ (lower left panel)
and $\Delta f_{\rm gas}  > 0.3$ (lower right panel). Solid lines show the corresponding
distributions for  the whole sample of simulated galaxies at 
available redshifts.  
}
\label{fig:histo_events}
\end{figure*}

We assumed that a galaxy has experienced an extreme event when one of the following
conditions is true:
$F_{\rm merger} > 0.3$ (Fig. \ref{fig:histo_events}, upper left panel), $F_{\rm inter}^{R_{\rm vir}}  > 0.6$ (Fig. \ref{fig:histo_events}, upper right panel),
$\Delta M_{\rm bar}/M_{\rm bar} < -0.3$ (Fig. \ref{fig:histo_events}, lower left panel)
and $\Delta f_{\rm gas}  > 0.3$ (Fig. \ref{fig:histo_events}, lower right panel).
$\Delta M_{\rm bar} / M_{\rm bar}$ is defined as the change in the baryonic mass of the galaxy
between two time steps of the simulation  if the time interval is not greater than 2 Gyr
and is normalised to the initial baryonic mass.
In these simulations, a decrease in the baryonic
mass of a galaxy ($\Delta M_{\rm bar} / M_{\rm bar}<0$) implies a significant loss of 
gas and, therefore, the condition $\Delta M_{\rm bar} / M_{\rm bar} < -0.3$ allows us to
select systems which have been subject to strong outflows events.  In fact, these
simulated galaxies would  have
lost at least 30\% of their baryonic mass during the last 2 Gyr.
$\Delta f_{\rm gas}$ is defined as the total variation in the gas fraction
of the galaxy with time.
In these simulations, the gas inside the galaxy tends to decrease because of the
star formation process or ejections by galactic winds. Hence, the gas
mass can only increase by infall of surrounding material and/or by mergers.
Therefore, our condition $\Delta f_{\rm gas}  > 0.3$  allows us to identify
galactic systems which have suffered important gas accretion.

We estimated the distribution
of the absolute variations of  $\log (V_{\rm rot} / V_{\rm circ})$  
and $\log (s_{\rm 1.0} / V_{\rm circ})$
of  galaxies which have suffered extreme events,
as shown in  Fig. \ref{fig:histo_events}.
For comparison, we calculated  
the corresponding distribution for the whole sample of simulated galaxies.
From the histograms, 
it is clear that the distribution of galaxies which have been
subject to extreme events are biased to larger values
of $|\Delta \log (V_{\rm rot} / V_{\rm circ})|$ than those of the whole galaxy sample, indicating that all these processes
tend to generate TFR outliers by disturbing the gas kinematics.
In particular, most events (from any of the mechanisms considered) induce
mean deviations of $\sim 0.1$ dex, with maximum offsets in the range [0.4,0.5]
dex.
We can also see that $|\Delta \log (V_{\rm rot} / V_{\rm circ}) | < 0.5$ 
but, if we use $s_{\rm 1.0}$ instead of $V_{\rm rot}$,
the distribution is narrower with  $|\Delta \log (s_{\rm 1.0} / V_{\rm circ}) | < 0.3$, approximately.
In the case of $|\Delta \log (s_{\rm 1.0} / V_{\rm circ})|$, disturbed galaxies tend to be biased
towards larger values 
when associated to mergers and interactions. However, 
the deviations of $s_{\rm 1.0} $ from $V_{\rm circ}$ are less significant 
in the case of  important inflows and outflows. This suggest that $s_{\rm 1.0} $ can account better for dispersion associated
to these later effects than to those which could be imprinted by more violent effects such as mergers and interactions.

In order to analyse if there is
any trend for some of the mechanisms to drive a positive or negative variation
of the rotation velocity relative to the circular one, we estimate the 
 percentage of galaxies with  $\Delta \log (V_{\rm rot} / V_{\rm circ}) < 0$ in each
of the subsamples. We found a weak trend for galaxies experiencing 
important mergers or  outflows  to have negative variations
($61 \%$ and $58\%$, respectively)
while strong interactions and gas inflows can drive either positive or
negative outliers on the TFR-plane ($49\%$ and $53\%$, respectively).

These results indicate that the joint
action of mergers, interactions, starbursts, outflows and gas accretion  can affect  the TFR   
producing  variations
in $\log (V_{\rm rot} / V_{\rm circ})$ as larger as $\sim 0.5$ dex, at least in   these
simulations.  
Hence, at a given stellar or baryonic mass, these events can produce a scatter in the TFR-plane larger than the mean level of velocity evolution
since $z \sim 3$  ($\sim 0.1$ dex). 
In fact, for any of the subsamples analysed in Fig. \ref{fig:histo_events},
while $\sim 25$\% of the galaxies have variations in $\log (V_{\rm rot} / V_{\rm circ})$
of $\sim 0.1$ dex, 55\% of the systems
exhibit values $>0.1$ dex,  
which might mask the evolution of the TFR in observational studies.
However, as we have seen, by combining $V_{\rm rot}$ and $\sigma$ in the definition of the kinematical indicator,
it is possible to generate a velocity scale more robust against disturbances of the gas kinematics
being able to reduce the scatter in the simulated sTFR and bTFR by a factor of $\sim 2$.

\section{Conclusions}
\label{sec:conclusions}

We studied the evolution of the gas kinematics of galaxies
by using numerical simulations
which include a physically motivated SN feedback.
We focused on the origin of the scatter of galaxies in the TFR-plane
and the connection with the formation histories of these systems.
In particular, 
we analysed the role of mergers and interactions on the determination
of gas kinematics and of the features of rotation curves during the assembly of galaxies.  
We extended the work by \citet{derossi2010},
who studied gas rotation-dominated systems, 
and explore the whole sample of galactic systems including also gas dispersion-dominated galaxies. 
It is worth mentioning that our simulated sample does not contain
realistic spiral-like galaxies as, in general, simulated systems
have dominant stellar spheroids. 
Nevertheless, all analysed galaxies have surviving gaseous discs 
which are found to be good tracers of the potential wells.

Our main results can be summarised as follows:

\begin{itemize}

\item Our simulations support the claim that  
$V_{\rm rot}$  is the best proxy for $V_{\rm circ}$ at $R_{\rm max}$ 
for galaxies of all morphology types. 
For gas rotation-dominated systems, 
$V_{\rm rot}$ is a good proxy of $V_{\rm circ}$ at all radii.
As the dispersion-dominated component increases,
$\sigma / V_{\rm rot}$  gets larger, specially in the outer parts of the systems. 
This leads to an increase of the scatter of the TFR
when considering systems with more disordered gas kinematics.
Nevertheless, for systems with large dispersion-dominated components, 
$V_{\rm rot}$ is still a good proxy for $V_{\rm circ}$ at $R_{\rm max}$.
In particular, for spheroidal dominated galaxies, we obtained 
a mean $\sigma / V_{\rm rot}$ of $\sim 0.7$ at $R_{\rm max}$
in good agreement with recent observations \citep{catinella2011}.

\item The use of the kinematical indicators  
$s_{0.5}$, 
$s_{1.0}$ and $S$ \citep{weiner2006,kassin2007,covington2010} in our galaxy sample, 
leads to a reduction of the scatter of the TFR in all cases.
Moreover, all of them, together with $V_{\rm rot}$, lead to the tightest relations if evaluated at $R_{\rm max}$. 
In particular, by comparing $s_{0.5}$ and $S$ with $V_{\rm circ}$, 
we obtained that the former underestimates $V_{\rm circ}$,
while the latter overproduces it. 
Although in the literature 
$s_{0.5}$ has been used more frequently, we
find that, in the case of these simulations, $s_{1.0}$ is a better kinematical
indicator as it does not only reduce the scatter of the TFR but it is also a good
proxy of $V_{\rm circ}$  at large radii ($r > 0.5 R_{\rm bar}$).
By restricting our study to subsamples with different baryonic $D/T$ or
different relative angular momentum orientation between the gas and stellar phases,
we obtained, on average, similar global trends.
Nevertheless, galaxies with misalignments between the gaseous and the stellar angular momenta 
or with lower baryonic $D/T$ present larger scatter around the mean behaviour.

\item 
Mergers and interactions within a $\Lambda -$CDM scenario can regulate the star formation process as
well as drive inflows and outflows of gas affecting the evolutionary tracks of galaxies 
on the TFR-plane and generating TFR outliers.
In particular, we obtained that merger-induced velocity disturbances in our
cosmological simulations are typically smaller ($<0.5$ dex) than
predicted by pre-prepared merger simulations of Milky-Way type galaxies \citep{covington2010}.
This is a consequence of the fact that the cosmological merger histories of the 
galaxies in the simulated mass range involved
typically more minor mergers 
than the major mergers commonly built in merger simulations.

\item 
We detected that mergers and interactions can generate either an increase or decrease
of $V_{\rm rot}/V_{\rm circ}$.  Nevertheless, statistically, we obtained a weak trend for galaxies subject to 
important mergers to be biased to negative variations ($\sim 61 \%$), while strong
interactions can generate either positive or negative changes with similar probability.
We also found that gas infall or outflows
can lead to TFR outliers.  
In particular, the statistical study shows a weak trend for outflows 
to produce negative variations ($\sim 58 \%$) while gas inflows can drive either positive or
negative changes.

\item
By comparing the distributions of the absolute variations of $\log (s_{\rm 1.0} / V_{\rm circ})$
and $\log (V_{\rm rot} / V_{\rm circ})$ during the evolution of all simulated galaxies,
we obtained that, for the former case, the scatter is reduced by a factor of $\sim 2$,
exhibiting a narrower distribution ($|\Delta \log (s_{\rm 1.0} / V_{\rm circ}) | < 0.3$).
In particular, mergers and interactions can be associated to a larger scatter in the TFR even 
when using $s_{\rm 1.0}$; conversely, the effects of inflows and outflows 
seem to be more efficiently accounted by this kinematical indicator.

\item According to our results, extreme physical events in simulated galaxies (e.g. mergers,
interactions, outflows and inflows) lead to a scatter on the TFR-plane
larger than the mean level of evolution of the TFR since $z \sim 3$.
In addition, by enhancing
the star formation rate,  mergers and interactions also affect the luminosity and $M/L$ ratio of the systems, which can
induce observational offsets from the TFR.
Therefore, much work is still needed from
the observational and theoretical point of views
to address these problems.

\end{itemize}

\begin{acknowledgements}
We thank the anonymous referee for his/her useful comments that largely
helped to improve this paper.
We thank Mario Abadi for useful comments.
We acknowledge support from the  PICT 32342 (2005),
PICT 245-Max Planck (2006) of ANCyT (Argentina), PIP 2009-112-200901-00305 of
CONICET (Argentina) and the L'oreal-Unesco-Conicet 2010 Prize.
Simulations were run in Fenix and HOPE clusters at IAFE.
\end{acknowledgements}


\begin{thebibliography}{64}
\bibliographystyle{aa}


\bibitem[Abadi et al.(2003)]{abadi2003} Abadi, M.~G., Navarro, 
J.~F., Steinmetz, M., \& Eke, V.~R.\ 2003, \apj, 591, 499

\bibitem[Amor{\'{\i}}n et 
al.(2009)]{amorin2009} Amor{\'{\i}}n, R., Aguerri, J.~A.~L., Mu{\~n}oz-Tu{\~n}{\'o}n, C., \& Cair{\'o}s, L.~M.\ 2009, \aap, 501, 75 


\bibitem[Atkinson et al.(2007)]{atkinson2007} Atkinson, N., 
Conselice, C.~J., \& Fox, N.\ 2007, arXiv:0712.1316

\bibitem[Avila-Reese et al.(1998)]{avila1998} Avila-Reese, V., 
Firmani, C., \& Hern{\'a}ndez, X.\ 1998, \apj, 505, 37 

\bibitem[Avila-Reese et al.(2008)]{avila2008} Avila-Reese, V., 
Zavala, J., Firmani, C., 
\& Hern{\'a}ndez-Toledo, H.~M.\ 2008, \aj, 136, 1340 

\bibitem[Barnes
\& Hernquist(1996)]{barnes1996} Barnes, J.~E., \& Hernquist, L.\ 1996, \apj, 471, 115


\bibitem[Barton et al.(2001)]{barton2001} Barton, E.~J., Geller, 
M.~J., Bromley, B.~C., van Zee, L., \& Kenyon, S.~J.\ 2001, \aj, 121, 625

\bibitem[Bell 
\& de Jong(2001)]{bell2001} Bell, E.~F., \& de Jong, R.~S.\ 2001, \apj, 550, 212 

\bibitem[B{\"o}hm et 
al.(2004)]{bohm2004} B{\"o}hm, A., et al.\ 2004, \aap, 420, 97

\bibitem[Catinella et al.(2012)]{catinella2011} Catinella, B., 
Kauffmann, G., Schiminovich, D., et al.\ 2012, \mnras, 420, 1959 

\bibitem[Conselice et al.(2005)]{concelise2005} Conselice, C.~J., 
Bundy, K., Ellis, R.~S., Brichmann, J., Vogt, N.~P., 
\& Phillips, A.~C.\ 2005, \apj, 628, 160 

\bibitem[Covington et al.(2010)]{covington2010} Covington, M.~D., et 
al.\ 2010, \apj, 710, 279


\bibitem[Cresci et al.(2009)]{cresci2009} Cresci, G., et al.\ 
2009, \apj, 697, 115 


\bibitem[de Rossi et 
al.(2010)]{derossi2010} de Rossi, M.~E., Tissera, P.~B., \& Pedrosa, S.~E.\ 2010, \aap, 519, A89


\bibitem[Flores et 
al.(2006)]{flores2006} Flores, H., Hammer, F., Puech, M., Amram, P., \& Balkowski, C.\ 2006, \aap, 455, 107 

\bibitem[Gnerucci et 
al.(2011)]{gnerucci2011} Gnerucci, A., et al.\ 2011, \aap, 528, A88

\bibitem[Guo et al.(2010)]{guo2010} Guo, Q., White, S., Li, C., 
\& Boylan-Kolchin, M.\ 2010, \mnras, 404, 1111 

\bibitem[Gurovich et al.(2010)]{gurovich2010} Gurovich, S., Freeman, 
K., Jerjen, H., Staveley-Smith, L., \& Puerari, I.\ 2010, \aj, 140, 663 


\bibitem[Kannappan 
\& Barton(2004)]{kannappan2004} Kannappan, S.~J., \& Barton, E.~J.\ 2004, \aj, 127, 2694

\bibitem[Kassin et al.(2007)]{kassin2007} Kassin, S.~A., et al.\ 
2007, \apjl, 660, L35

\bibitem[McCarthy et al.(2012)]{mccarthy2012} 
McCarthy, I.~G., Schaye, J., Font, A.~S., Theuns, T., Frenk, C.~S.,
Crain, R.~A., \& Dalla Vecchia, C.\ 2012, arXiv:1204.5195

\bibitem[McGaugh et al.(2000)]{mcgaugh2000} McGaugh, S.~S., 
Schombert, J.~M., Bothun, G.~D., 
\& de Blok, W.~J.~G.\ 2000, \apjl, 533, L99 

\bibitem[Meyer et al.(2008)]{meyer2008} Meyer, M.~J., Zwaan, 
M.~A., Webster, R.~L., Schneider, S., 
\& Staveley-Smith, L.\ 2008, \mnras, 391, 1712 

\bibitem[Mihos
\& Hernquist(1996)]{mihos1996} Mihos, J.~C., \& Hernquist, L.\ 1996, \apj, 464, 641

\bibitem[Miller et al.(2011)]{miller2011} Miller, S.~H., Bundy, 
K., Sullivan, M., Ellis, R.~S., \& Treu, T.\ 2011, \apj, 741, 115 

\bibitem[Mo et al.(1998)]{mo1998} Mo, H.~J., Mao, S., 
\& White, S.~D.~M.\ 1998, \mnras, 295, 319 

\bibitem[Mosconi et al.(2001)]{mosconi2001} Mosconi, M.~B., 
Tissera, P.~B., Lambas, D.~G., \& Cora, S.~A.\ 2001, \mnras, 325, 34 

\bibitem[Moster et al.(2010)]{moster2010} Moster, B.~P., 
Somerville, R.~S., Maulbetsch, C., van den Bosch, F.~C., Macci{\`o}, A.~V., 
Naab, T., \& Oser, L.\ 2010, \apj, 710, 903 


\bibitem[Nakamura et al.(2006)]{nakamura2006} Nakamura, O., 
Arag{\'o}n-Salamanca, A., Milvang-Jensen, B., Arimoto, N., Ikuta, C., 
\& Bamford, S.~P.\ 2006, \mnras, 366, 144


\bibitem[Pedrosa et 
al.(2008)]{pedrosa2008} Pedrosa, S., Tissera, P.~B., Fuentes-Carrera, I., \& Mendes de Oliveira, C.\ 2008, \aap, 484, 299

\bibitem[Persic
\& Salucci(1995)]{persic1995} Persic, M., \& Salucci, P.\ 1995, \apjs, 99, 501


\bibitem[Pizagno et al.(2007)]{pizagno2007} Pizagno, J., et al.\ 
2007, \aj, 134, 945

\bibitem[Portinari 
\& Sommer-Larsen(2007)]{portinari2007} Portinari, L., \& Sommer-Larsen, J.\ 2007, \mnras, 375, 913

\bibitem[Puech et 
al.(2008)]{puech2008} Puech, M., et al.\ 2008, \aap, 484, 173

\bibitem[Reyes et al.(2011)]{reyes2011} Reyes, R.,
Mandelbaum, R., Gunn, J.~E., Pizagno, J. \& Lackner, C.N.\ 2011, \mnras, 417, 2347

\bibitem[Scannapieco et al.(2005)]{scannapieco2005} Scannapieco, C., 
Tissera, P.~B., White, S.~D.~M., \& Springel, V.\ 2005, \mnras, 364, 552 

\bibitem[Scannapieco et al.(2006)]{scannapieco2006} Scannapieco, C., 
Tissera, P.~B., White, S.~D.~M., \& Springel, V.\ 2006, \mnras, 371, 1125 

\bibitem[Scannapieco et al.(2008)]{scannapieco2008} Scannapieco, C., 
Tissera, P.~B., White, S.~D.~M., \& Springel, V.\ 2008, \mnras, 389, 1137

\bibitem[Scannapieco et al.(2009)]{scan09} Scannapieco, C., 
White, S.~D.~M., Springel, V., \& Tissera, P.~B.\ 2009, \mnras, 396, 696

\bibitem[Simard 
\& Pritchet(1998)]{simard1998} Simard, L., \& Pritchet, C.~J.\ 1998, \apj, 505, 96

\bibitem[Snaith al.(2012)]{snaith2012} Snaith, O, Gibson, B. et al. \ 2012, \mnras, in press

\bibitem[Springel et al.(2001)]{springel2001} Springel, V., White, 
S.~D.~M., Tormen, G., \& Kauffmann, G.\ 2001, \mnras, 328, 726 

\bibitem[Springel 
\& Hernquist(2003)]{springel2003} Springel, V., \& Hernquist, L.\ 2003, \mnras, 339, 289 

\bibitem[Springel(2005)]{springel2005} Springel, V.\ 2005, \mnras, 
364, 1105 

\bibitem[Stewart et al.(2009)]{stewart2009} Stewart, K.~R., 
Bullock, J.~S., Wechsler, R.~H., \& Maller, A.~H.\ 2009, \apj, 702, 307 

\bibitem[Thielemann et al.(1993)]{thielemann1993} Thielemann, F.-K., 
Nomoto, K., 
\& Hashimoto, M.\ 1993, Origin and Evolution of the Elements, 297 


\bibitem[Tissera(2000)]{tissera2000} Tissera, P.~B.\ 2000, \apj,
534, 636

\bibitem[Tissera et al.(2012)]{tiss2011} Tissera, P.~B., White, 
S.~D.~M., \& Scannapieco, C.\ 2012, \mnras, 420, 255 

\bibitem[Torres-Flores et al.(2011)]{torres2011} Torres-Flores, 
S., Epinat, B., Amram, P., Plana, H., 
\& Mendes de Oliveira, C.\ 2011, \mnras, 416, 1936 

\bibitem[Tully 
\& Fisher(1977)]{tully1977} Tully, R.~B., \& Fisher, J.~R.\ 1977, \aap, 54, 661 

\bibitem[Tully 
\& Fouque(1985)]{tully1985} Tully, R.~B., \& Fouque, P.\ 1985, \apjs, 58, 67

\bibitem[Vogt et al.(1996)]{vogt1996} Vogt, N.~P., Forbes, 
D.~A., Phillips, A.~C., Gronwall, C., Faber, S.~M., Illingworth, G.~D., 
\& Koo, D.~C.\ 1996, \apjl, 465, L15

\bibitem[Vogt et al.(1997)]{vogt1997} Vogt, N.~P., et al.\ 1997, 
\apjl, 479, L121

\bibitem[Weiner et al.(2006)]{weiner2006} Weiner, B.~J., et al.\ 
2006, \apj, 653, 1027

\bibitem[Woosley 
\& Weaver(1995)]{woosley1995} Woosley, S.~E., \& Weaver, T.~A.\ 1995, \apjs, 101, 181 

\bibitem[Yegorova
\& Salucci(2007)]{yegorova2007} Yegorova, I.~A., \& Salucci, P.\ 2007, \mnras, 377, 507


\end{thebibliography}
\end{document}